\documentclass[journal]{IEEEtran}
\usepackage{amsmath,amsfonts}
\usepackage{algorithmic}
\usepackage{algorithm}
\usepackage{array}
\usepackage[caption=false,font=normalsize,labelfont=sf,textfont=sf]{subfig}
\usepackage{textcomp}
\usepackage{stfloats}
\usepackage{url}
\usepackage{verbatim}
\usepackage{graphicx}

\usepackage{longtable}

\usepackage{cite}
\usepackage{multirow}
\usepackage{makecell}

\usepackage{booktabs}
\usepackage[table,x11names]{xcolor}
\definecolor{lightblue}{RGB}{173, 216, 230}
\definecolor{green}{RGB}{144,238,144} 

\begin{document}

\title{Voice Cloning: Comprehensive Survey}

\author{Hussam~Azzuni, and
        Abdulmotaleb~El Saddik,~\IEEEmembership{Fellow,~IEEE,}
\thanks{Hussam Azzuni performed this work during his stay with the Computer Vision Department, MBZUAI, UAE}
\thanks{Abdulmotaleb El Saddik is with the University of Ottawa, Canada}


}



\maketitle

\begin{abstract}
Voice Cloning has rapidly advanced in today's digital world, with many researchers and corporations working to improve these algorithms for various applications. This article aims to establish a standardized terminology for voice cloning and explore its different variations. It will cover speaker adaptation as the fundamental concept and then delve deeper into topics such as few-shot, zero-shot, and multilingual TTS within that context. Finally, we will explore the evaluation metrics commonly used in voice cloning research and related datasets. This survey compiles the available voice cloning algorithms to encourage research toward its generation and detection to limit its misuse.

\end{abstract}


\section{Introduction}

\IEEEPARstart{V}{oice} Cloning is the ability to replicate a person's voice. Advancing these algorithms relies on enhancing the performance of Text-to-Speech (TTS) systems in various areas, including speech quality, naturalness, prosody, and timbre, ensuring the produced voice closely resembles the target speaker. The development of voice cloning can revolutionize many fields as we shift toward a more digital world. These applications include voice-over, personalized virtual assistants, and Human-Computer Interaction (HCI). 

\begin{figure}
\centering
  \includegraphics[width=0.95\linewidth]{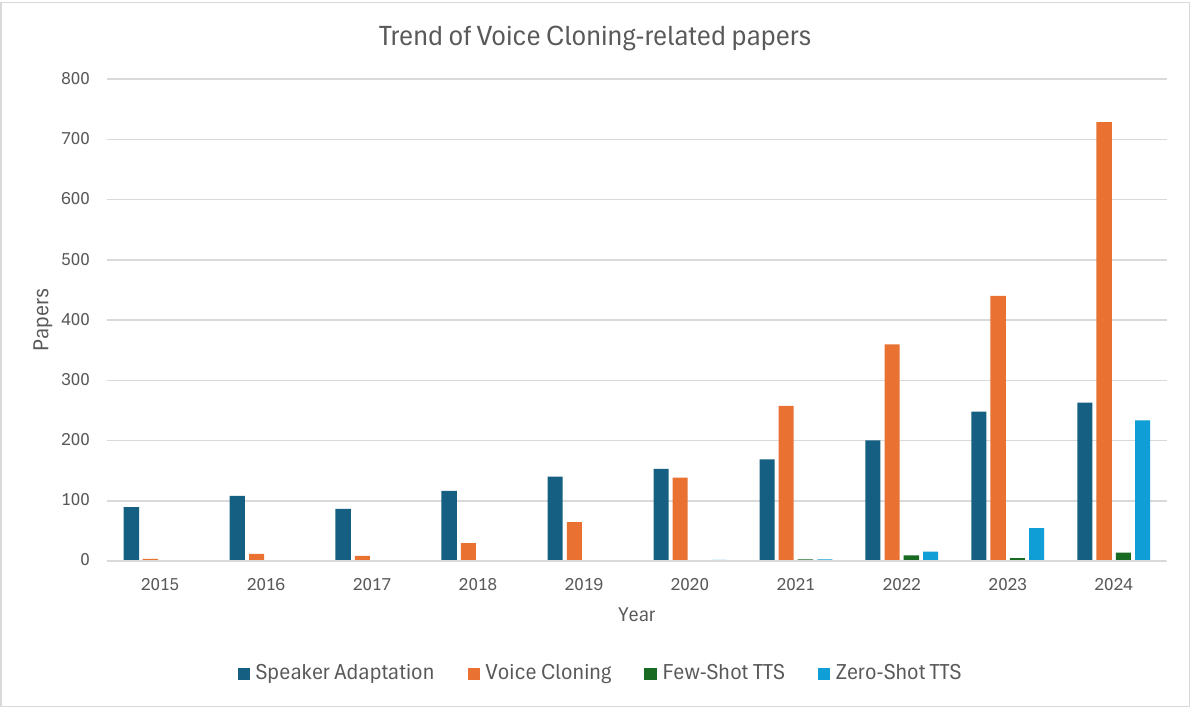}
\caption{Bar graph showing the trend in domains related to speaker adaptation.}
\label{VoiceCloning_Trend}  
\end{figure}

Voice cloning has been progressing rapidly, as demonstrated in Fig. \ref{VoiceCloning_Trend}; thus, establishing a comprehensive survey is crucial for further research in its generation and detection. The first step is reviewing previous literature that discussed deepfake generation and detection summarized in Table \ref{Previous_Survey_Papers}. We explore these works to gain insights before building on more recent methods in the domain. Tan et al. \cite{Tan_Survey_arxiv2021} provided a comprehensive analysis of neural speech synthesis while delving slightly into more advanced topics, including speaker adaptation. Kadam et al. \cite{Kadam_Survey_EAI2021} surveyed methods from voice cloning and lip synchronization, including techniques such as concatenative TTS and parametric TTS. Masood et al. \cite{Masood_Survey_AI2023} explored deepfake generation and detection for visual and audio modalities. This survey includes visual manipulation such as face swapping, puppet mastery, lip-syncing, entire face synthesis, and facial attribute manipulation. It also focuses on TTS and voice conversion (VC) for audio deepfakes. Similarly, Khanjani et al. \cite{Khanjani_Survey_FBD2023} explored audio, text, video, and image deepfakes while delving deeper into audio deepfake generation and detection, including both TTS and VC. More recently, Ramu et al. \cite{Ramu_Survey_WiSPNet2024} explored voice cloning in the context of multilingual video dubbing. Our survey differs from the previous papers as we will investigate and provide an in-depth study of audio deepfake generation, specifically in TTS systems, commonly called voice cloning. This document includes establishing a standardized terminology while going into more advanced topics, including few-shot, zero-shot, and multilingual TTS.

Voice cloning definition has varied across research. It differs from speech synthesis as it aims to preserve the speaker's characteristics. However, the amount of data required is in question. Some research refers to it similarly to speaker adaptation with no constraint on data availability \cite{Masood_Survey_AI2023, Zhao_ICCNEA2020, Luong_Nautilus_TASLP2020}. On the other hand, others state that voice cloning requires the usage of limited amount of data to replicate the target speaker's voice \cite{Arik_NIPS2018, Cong_InterSpeech2020, Shaheen_InterSpeech2023, Mandeel_SpeD2023, Huang_MetaTTS_TASLP2022, Huang_ICASSP2023, Nakai_APSIPAASC2022}. Nautilus \cite{Luong_Nautilus_TASLP2020} takes another approach in defining voice cloning, as the authors state that voice cloning is a big umbrella of any technology that can generate speech imitating a speaker's characteristic. Cross-lingual voice cloning \cite{Zhang_InterSpeech2019} differs from the previous definitions as it adapts known speakers to unknown languages, which is not within the scope of this survey. For this work, we propose establishing a standard definition for voice cloning as an extension of TTS speaker adaptation. It is essential to note that Table \ref{Abreviations} will cover all the abbreviations within this document.

\begin{enumerate}
    \item \underline{Voice Cloning:} Replicating a specific person's voice using a TTS system.
    \item \underline{Speaker Adaptation:} Fine-tuning a TTS model to replicate a specific user's voice using limited data.
    \item \underline{Few-shot Voice Cloning:} This follows the central concept of speaker adaptation. However, the difference is the amount of data required. Thus, the reference audio can range from a few seconds to a maximum of 5 minutes, which is decided based on previous work, and anything more is challenging to obtain in real-life scenarios.
    \item \underline{Zero-shot Voice Cloning:} This fundamentally differs from the previous approaches as you do not need to fine-tune the TTS model. A specialized module, such as a speaker encoder, is required to use a short audio clip to generate speech with voice characteristics similar to the reference waveform.
\end{enumerate}

We follow a multi-stage process to include papers in this survey. First, we review the previous literature on TTS speaker adaptation and voice cloning and select the articles aligned with our scope. Second, we perform a comprehensive search for relevant English methodology papers using keywords such as "voice cloning," "few-shot voice cloning," "zero-shot voice cloning," and "speaker adaptation" to find relevant papers. We exclude voice conversion, singing voice synthesis (SVS), and speech enhancement (SE) papers. Additionally, we exclude 2024 non-peer-reviewed documents, as they may still be works in progress unless they have already had a significant impact.

This survey will comprehensively explore the current status of voice cloning, including speaker adaptation, few-shot, zero-shot, and multilingual voice cloning. Establishing a voice cloning reference paper is crucial for further development and exploring audio deepfake generation and its detection to limit its misuse in real-world scenarios. Section \ref{Section2} provides a taxonomy of TTS system components. Understanding the main components is crucial for developing the basic knowledge of voice cloning within various models. Section \ref{Section3} explains background information regarding the most common TTS algorithms as they are the basis of much recent work. In addition, information regarding pre-training and training algorithms is needed for an in-depth understanding of some contributions. Section \ref{Section4} explores voice cloning variants with additional subcategories based on researchers' contributions. Section \ref{Section5} goes through speech-related datasets and explores the evaluation metrics used for TTS systems. Section \ref{Section6} goes through potential applications for voice cloning across various sectors and certain misuse cases to familiarize the reader with the dangers of voice cloning. Finally, Section \ref{Section7} will discuss future directions to ensure growth in voice cloning.

\begin{table*}[t]
\centering
\caption{Relevant Literature review.}
\resizebox{18cm}{!}{
\begin{tabular}{ccccc}
\toprule
\textbf{Paper} & \textbf{Year} & \textbf{Conference/Journal} & \textbf{Scope}\\ \midrule

\multicolumn{1}{l|}{Ref. \cite{Ramu_Survey_WiSPNet2024}} &  
\multicolumn{1}{c|}{2024} &  
\multicolumn{1}{c|}{International Conference on Wireless Communication Signal Processing and Networking} &  
\multicolumn{1}{c}{Multilingual video dubbing, focusing on voice cloning and machine translation}\\

\midrule

\multicolumn{1}{l|}{Ref. \cite{Khanjani_Survey_FBD2023}} &  
\multicolumn{1}{c|}{2023} &  
\multicolumn{1}{c|}{Frontiers in Big Data} &  
\multicolumn{1}{c}{Audio deepfake generation and detection}\\

\midrule

\multicolumn{1}{l|}{Ref. \cite{Masood_Survey_AI2023}} &  
\multicolumn{1}{c|}{2023} &  
\multicolumn{1}{c|}{Applied Intelligence} &  
\multicolumn{1}{c}{Audio and video deepfake generation and detection}\\
 
\midrule

\multicolumn{1}{l|}{Ref. \cite{Kadam_Survey_EAI2021}} &  
\multicolumn{1}{c|}{2021} &  
\multicolumn{1}{c|}{EAI Endorsed Transactions on Creative Technologies} &  
\multicolumn{1}{c}{Voice cloning and lip synchronization}\\

\midrule

\multicolumn{1}{l|}{Ref. \cite{Tan_Survey_arxiv2021}} &  
\multicolumn{1}{c|}{2021} &  
\multicolumn{1}{c|}{Arxiv} &  
\multicolumn{1}{c}{Neural TTS, including components and advanced topics}\\

\bottomrule

\end{tabular}}
\label{Previous_Survey_Papers}
\end{table*}

\begin{table}[t]
\centering
\caption{Table of Common Abbreviations.}
\resizebox{8.0cm}{!}{
\begin{tabular}{cc||cc}
\toprule
\textbf{Abbreviation} & \textbf{Full Name} & \textbf{Abbreviation} & \textbf{Full Name}  \\ \midrule

\rowcolor{lightgray} 
\multicolumn{4}{c}{TTS systems} \\
\midrule

\multicolumn{1}{l|}{FastSpeech 1/2} &  
\multicolumn{1}{l||}{FS/FS2}
&
\multicolumn{1}{l|}{StyleTTS \& StyleTTS2} &  
\multicolumn{1}{l}{STTS/STTS2} \\

\midrule

\multicolumn{1}{l|}{NaturalSpeech 2/3} &  
\multicolumn{1}{l||}{NS2/NS3}
&
\multicolumn{1}{l|}{Guided-TTS \& Guided-TTS 2} &  
\multicolumn{1}{l}{GTTS/GTTS2}  \\

\midrule

\multicolumn{1}{l|}{MegaTTS \& MegaTTS2} &  
\multicolumn{1}{l||}{MTTS/MTTS2}
&
\multicolumn{1}{l|}{HierSpeech \& HierSpeech++} &  
\multicolumn{1}{l}{HS/HS++}  \\

\midrule

\multicolumn{1}{l|}{YourTTS} &  
\multicolumn{1}{l||}{YTTS}
&
\multicolumn{1}{l|}{Content-Dependent Fine-Grained Speaker Embedding} &  
\multicolumn{1}{l}{CDFSE} \\

\midrule

\multicolumn{1}{l|}{StyleSpeech \& Meta-StyleSpeech} &  
\multicolumn{1}{l||}{SS/MSS}
&
\multicolumn{1}{l|}{VoiceBox} &  
\multicolumn{1}{l}{VB}  \\

\midrule

\rowcolor{lightgray} 
\multicolumn{4}{c}{Fundamental Neural Network Concepts} \\
\midrule

\multicolumn{1}{l|}{Variational Autoencoder} &  
\multicolumn{1}{l||}{VAE} 
&
\multicolumn{1}{l|}{Convolutional Neural Networks} &  
\multicolumn{1}{l}{CNN}  \\

\midrule

\multicolumn{1}{l|}{Deep Neural Network} &  
\multicolumn{1}{l||}{DNN}  
&
\multicolumn{1}{l|}{Recurrent Neural Network} &  
\multicolumn{1}{l}{RNN}  \\

\midrule

\multicolumn{1}{l|}{Bidirectional Long Short-Term memory} &  
\multicolumn{1}{l||}{Bi-LSTM}  
&
\multicolumn{1}{l|}{Generative Adversarial Network} &  
\multicolumn{1}{l}{GAN}  \\

\midrule

\multicolumn{1}{l|}{Sequence-to-Sequence} &  
\multicolumn{1}{l||}{Seq2Seq}
&
\multicolumn{1}{l|}{Denoising Diffusion Probabilistic Models} &  
\multicolumn{1}{l}{DDPM} \\

\midrule

\multicolumn{1}{l|}{Self-Supervised Learning} &  
\multicolumn{1}{l||}{SSL}  
&
\multicolumn{1}{l|}{Model-Agnostic Meta-Learning} &  
\multicolumn{1}{l}{MAML}  \\

\midrule

\multicolumn{1}{l|}{Language-Agnostic Meta-Learning} &  
\multicolumn{1}{l||}{LAML} 
&
\multicolumn{1}{l|}{Autoregressive \& Non-Autoregressive} & 
\multicolumn{1}{l}{AR/NAR}  \\

\midrule

\rowcolor{lightgray} 
\multicolumn{4}{c}{Language Preprocessing} \\
\midrule

\multicolumn{1}{l|}{Byte-Pair encoding} &  
\multicolumn{1}{l||}{BPE}
&
\multicolumn{1}{l|}{Grapheme to Phoneme} &  
\multicolumn{1}{l}{G2P}  \\

\midrule

\multicolumn{1}{l|}{International Phonetic Alphabet} &  
\multicolumn{1}{l||}{IPA}
&
\multicolumn{1}{l|}{-} &  
\multicolumn{1}{l}{-}  \\

\midrule

\rowcolor{lightgray} 
\multicolumn{4}{c}{Speech Representation} \\
\midrule

\multicolumn{1}{l|}{Fundamental frequency} &  
\multicolumn{1}{l||}{f0}
&
\multicolumn{1}{l|}{Vector-Quantization VAE} &  
\multicolumn{1}{l}{VQ-VAE}  \\

\midrule

\multicolumn{1}{l|}{Phoneme-Posteriorgram} &  
\multicolumn{1}{l||}{PPG}
&
\multicolumn{1}{l|}{Five-time sampling} &  
\multicolumn{1}{l}{FTS}  \\

\midrule

\multicolumn{1}{l|}{Lookup Table} &  
\multicolumn{1}{l||}{LUT}
&
\multicolumn{1}{l|}{Learnable Dictionary Encoding} &  
\multicolumn{1}{l}{LDE}  \\

\midrule

\rowcolor{lightgray} 
\multicolumn{4}{c}{TTS and speech-related applications} \\

\midrule

\multicolumn{1}{l|}{Human-Computer Interaction} & 
\multicolumn{1}{l||}{HCI}
&
\multicolumn{1}{l|}{Singing Voice Synthesis} &  
\multicolumn{1}{l}{SVS}  \\

\midrule

\multicolumn{1}{l|}{Text to Speech} &  
\multicolumn{1}{l||}{TTS}  
&
\multicolumn{1}{l|}{Few-Shot TTS \& Zero-Shot TTS} &  
\multicolumn{1}{l}{FS-TTS/ZS-TTS} \\

\midrule

\multicolumn{1}{l|}{Language Identification} &  
\multicolumn{1}{l||}{LID}  
&
\multicolumn{1}{l|}{Language Understanding} &  
\multicolumn{1}{l}{LU} \\

\midrule

\multicolumn{1}{l|}{Machine Translation} &  
\multicolumn{1}{l||}{MT}
&
\multicolumn{1}{l|}{Speech-to-Speech Translation} &  
\multicolumn{1}{l}{S2ST}  \\

\midrule

\multicolumn{1}{l|}{Automatic Speech Recognition} &  
\multicolumn{1}{l||}{ASR}
&
\multicolumn{1}{l|}{Automatic Phoneme Recognition} &  
\multicolumn{1}{l}{APR}  \\

\midrule

\multicolumn{1}{l|}{Speaker Verification} &  
\multicolumn{1}{l||}{SV}
&
\multicolumn{1}{l|}{Speaker Identification} &  
\multicolumn{1}{l}{SI}  \\

\midrule

\multicolumn{1}{l|}{Voice Conversion} &  
\multicolumn{1}{l||}{VC}
&
\multicolumn{1}{l|}{Emotional Voice Conversion} &  
\multicolumn{1}{l}{EVC}  \\

\midrule

\multicolumn{1}{l|}{Speech Emotion Recognition} &  
\multicolumn{1}{l||}{SER}
&
\multicolumn{1}{l|}{Speech Enhancement} &  
\multicolumn{1}{l}{SE}  \\

\midrule

\rowcolor{lightgray} 
\multicolumn{4}{c}{Speaker \& Style Modeling} \\
\midrule

\multicolumn{1}{l|}{Global Style Token} &  
\multicolumn{1}{l||}{GST}
&
\multicolumn{1}{l|}{Style-adaptive Layer Normalization} &  
\multicolumn{1}{l}{SALN}  \\

\midrule

\multicolumn{1}{l|}{Adaptive Instance Normalization} &  
\multicolumn{1}{l||}{AdaIN}
&
\multicolumn{1}{l|}{Adaptive Layer Normalization} &  
\multicolumn{1}{l}{AdaLN}  \\

\midrule

\multicolumn{1}{l|}{Conditional Layer Normalization} &  
\multicolumn{1}{l||}{CLN}
&
\multicolumn{1}{l|}{-} &  
\multicolumn{1}{l}{-}  \\

\midrule

\rowcolor{lightgray} 
\multicolumn{4}{c}{Language models in TTS systems} \\
\midrule

\multicolumn{1}{l|}{Large Language Model} &  
\multicolumn{1}{l||}{LLM}
&
\multicolumn{1}{l|}{Neural Codec Language Model} &  
\multicolumn{1}{l}{NCLM}  \\

\midrule

\rowcolor{lightgray}
\multicolumn{4}{c}{Evaluation} \\
\midrule

\multicolumn{1}{l|}{Mean Opinion Score} &  
\multicolumn{1}{l||}{MOS}  
&
\multicolumn{1}{l|}{Comparative MOS} &  
\multicolumn{1}{l}{CMOS}  \\

\midrule

\multicolumn{1}{l|}{Speaker Verification Equal Error Rate} &  
\multicolumn{1}{l||}{SV-EER}  
&
\multicolumn{1}{l|}{Word/Character Error Rate} &  
\multicolumn{1}{l}{WER/CER}  \\

\midrule

\multicolumn{1}{l|}{Real-Time Factor} &  
\multicolumn{1}{l||}{RTF}  
&
\multicolumn{1}{l|}{Speaker Embedding Cosine Similarity} &  
\multicolumn{1}{l}{SECS}  \\

\midrule

\multicolumn{1}{l|}{Gross Pitch Error} &  
\multicolumn{1}{l||}{GPE}
&
\multicolumn{1}{l|}{Voicing Decision Error} &  
\multicolumn{1}{l}{VDE}  \\

\midrule

\multicolumn{1}{l|}{F0 Frame Error} &  
\multicolumn{1}{l||}{FFE} 
&
\multicolumn{1}{l|}{Mel-Cepstral Distortion} &  
\multicolumn{1}{l}{MCD}  \\

\midrule

\multicolumn{1}{l|}{Mel-Frequency Cepstral Coefficient} &  
\multicolumn{1}{l||}{MFCC} &
\multicolumn{1}{l|}{-} &  
\multicolumn{1}{l}{-}  \\

\bottomrule

\end{tabular}}
\label{Abreviations}
\end{table}

\section{Text-to-Speech Taxonomy}
\label{Section2}

Voice Cloning is the process of adapting a TTS system to unseen speakers. Understanding its various categories is crucial in building on previous knowledge and making novel contributions. These systems follow a modular approach, starting from a text/phoneme until an audio waveform. TTS systems use various technologies such as Convolutional Neural Networks (CNN), Variational autoencoders (VAE), Generative Adversarial Networks (GAN), Flow-based generative models, Diffusion Models, and more recently, Codec-based networks. Here are the main three components.

\begin{enumerate}
    \item \underline{Text-Analysis:} This is the first module of a TTS system, known as the front-end, which generates linguistic features from a text input. This process includes text normalization, word segmentation if applicable, part-of-speech tagging, prosody prediction, and grapheme-to-phoneme (G2P) conversion \cite{Tan_Survey_arxiv2021}. 
        
    \item \underline{Acoustic Model:} This module, called the parameter prediction module, generates acoustic features from linguistic features or text/phonemes input, focusing on voice characteristics like prosody and timbre. Prosody modeling contains elements such as the speech's rhythm, intonation, and stress to enable expressive speech \cite{Chien_SLT2021}, evaluated through fundamental frequency (f0), duration, and energy. On the other hand, timbre represents the unique qualities of every person's voice.
                
    \item \underline{Vocoder:} This module takes acoustic or linguistic features and uses them to generate an audio waveform.
    
\end{enumerate}

\section{Relevant Background information}
\label{Section3}

This section will discuss the basic principles of various topics to ensure the reader has the necessary knowledge to better understand authors' contributions to voice cloning and its variants. These topics include common TTS algorithms, pre-training algorithms, and training paradigms.

\begin{figure*}
\centering
  \includegraphics[width=0.96\linewidth]{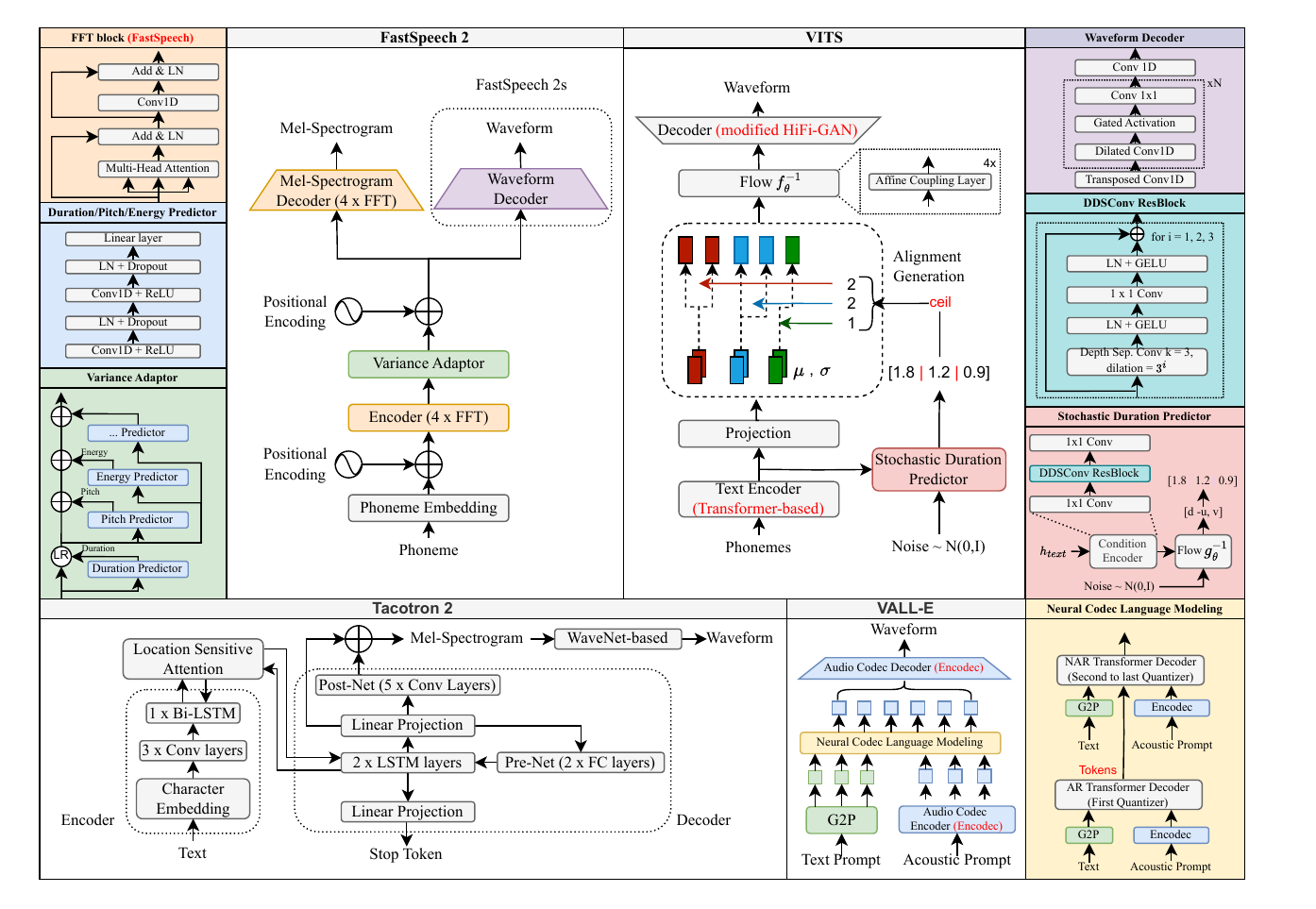}
\caption{Common TTS algorithms, adapted from Tacotron 2 \cite{Shen_Tacotron2_ICASSP2018}, FastSpeech 2 \cite{Ren_FastSpeech2_ICLR2021}, VITS \cite{Kim_VITS_ICML2021}, and VALL-E \cite{Wang_VALLE_arxiv2023}.}
\label{CommonTTS}  
\end{figure*}

\subsection{Common TTS algorithms}
TTS algorithms can be categorized into various architectural classes. Fig. \ref{CommonTTS} shows the most common TTS models. Tacotron 2 \cite{Shen_Tacotron2_ICASSP2018} builds upon Tacotron \cite{Wang_Tacotron_InterSpeech2017} by introducing key modifications to ensure smoother training and better speech quality without relying on feature engineering. It is a sequence-to-sequence (Seq2Seq) model with an encoder-decoder architecture and an attention mechanism to map a text input sequence to a mel spectrogram sequence during acoustic modeling. Unlike Tacotron, which uses the Griffin-Lim algorithm \cite{Griffin_Vocoder_1984}, Tacotron 2 replaces it with an independently trained, modified WaveNet vocoder \cite{Dieleman_WaveNet_arxiv2016}, leading to an improved synthesized speech quality. Similarly, FastSpeech 1/2 \cite{Ren_FastSpeech_NIPS2019, Ren_FastSpeech2_ICLR2021} is a transformer-based Seq2Seq model that uses feed-forward network instead of an encoder-decoder architecture. This modification speeds up the generation process by adopting a non-autoregressive (NAR) generation methodology compared to the autoregressive (AR) Tacotron 2. FastSpeech 2 builds on top of its predecessor by removing the distillation process and training directly on ground-truth mel spectrogram and the introduction of conditional inputs to handle the one-to-many mapping problem; this problem occurs as words are spoken in multiple ways from the same text. It also introduces FastSpeech 2s, the first attempt to generate a speech waveform in parallel through a GAN-based waveform decoder for end-to-end speech generation. VITS \cite{Kim_VITS_ICML2021} is an end-to-end TTS architecture that utilizes VAE and flow-based modeling to unify acoustic and vocoder representations for speech synthesis. It enhances the speech's expressiveness by applying normalizing flows to the conditional prior distribution and improves speech quality using waveform-based adversarial training. It also tackles the one-to-many relationship problem by introducing a stochastic duration predictor, enabling speech generation with diverse rhythms. With the emergence of Large Language Models (LLMs), TTS algorithms started integrating language modeling to enhance performance. VALL-E \cite{Wang_VALLE_arxiv2023} is a Neural Codec Language Model (NCLM) that regards TTS as a conditional language modeling task instead of a continuous signal regression task. Instead of obtaining an intermediate mel spectrogram, it uses discrete units based on phoneme and acoustic prompts. This model is capable of zero-shot TTS (ZS-TTS) using an acoustic prompt input to generate a waveform maintaining the speaker's emotion and voice characteristics. It consists of 4 components: a G2P module, an Audio Codec encoder and decoder following EnCodec \cite{Defossez_encodec_TML2023}, and an NCLM. The model is trained through in-context learning, similar to GPT-3 \cite{Brown_GPT3_NIPS2020}, enabling prompt-based approaches for ZS-TTS.

\subsection{Speech Pre-training Algorithms}
Understanding speech pre-training algorithms is essential, as some works utilize a pre-trained network for feature extraction. They can be categorized based on input, output, and downstream tasks, as shown in Table \ref{SpeechPretraining}. Wav2Vec 2.0 \cite{Baevski_Wav2Vec_NIPS2020}, Hubert \cite{Hsu_Hubert_TASLP2021}, and SLAM \cite{Bapna_SLAM_arxiv2021} all outputs discrete speech units for speech understanding as observed from their downstream tasks. On the other hand, Masked Acoustic Modeling (MAM) \cite{Chen_MAM_arxiv2020} and FAT-MLM \cite{Zheng_FATMLM_ICML2021} outputs reconstructed spectrogram as part of their training. More recently, A$^3$T \cite{Bai_A3T_ICML2022} is a pre-training framework for direct application for Text-to-Speech systems as an alternative to Wav2Vec 2.0 and Hubert.

\begin{figure*}
\centering
  \includegraphics[width=0.96\linewidth]{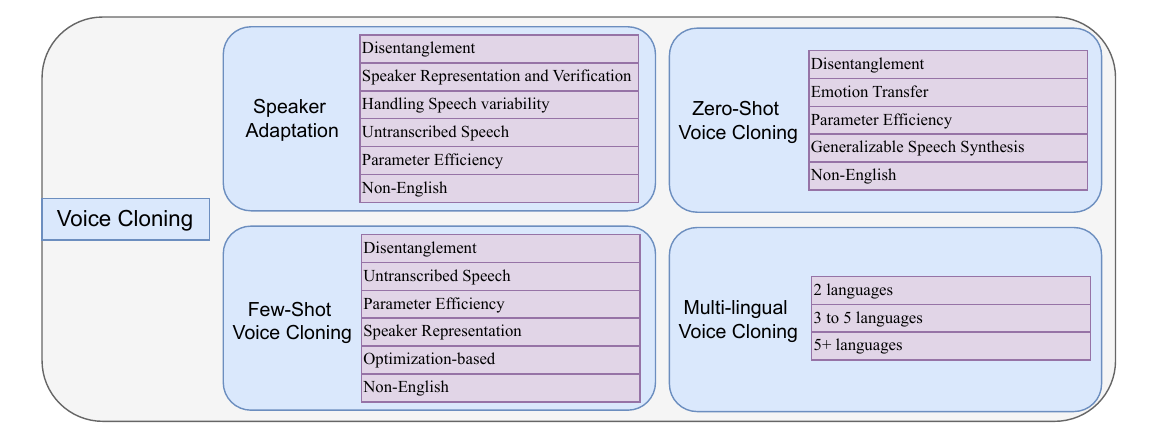}
\caption{Overview of the Survey paper.}
\label{VoiceCloning_SurveyOverview}  
\end{figure*}

\begin{table}[t]
\caption{Speech Pre-training algorithms adapted from \cite{Bai_A3T_ICML2022}.}
\centering
\resizebox{8.5cm}{!}{
\begin{tabular}{ccccc}
\toprule
\textbf{Model} & \textbf{Year} & \textbf{Input} & \textbf{Output}& \textbf{Downstream tasks} \\
\midrule
\multicolumn{1}{l|}{Wav2Vec 2.0 \cite{Baevski_Wav2Vec_NIPS2020}} & 
\multicolumn{1}{l}{2020} & 
\multicolumn{1}{|l|}{Speech} & 
\multicolumn{1}{l|}{Discrete units} & 
\multicolumn{1}{l}{ASR, APR} \\

\midrule

\multicolumn{1}{l|}{Hubert \cite{Hsu_Hubert_TASLP2021}} & 
\multicolumn{1}{l}{2021} & 
\multicolumn{1}{|l|}{Speech} & 
\multicolumn{1}{l|}{Discrete Units} & 
\multicolumn{1}{l}{ASR} \\

\midrule

\multicolumn{1}{l|}{SLAM \cite{Bapna_SLAM_arxiv2021}} & 
\multicolumn{1}{l}{2021} & 
\multicolumn{1}{|l|}{Speech, $<$Text-Speech$>$} & 
\multicolumn{1}{l|}{Discrete Units} & 
\multicolumn{1}{l}{ASR, ST, LU, TN} \\

\midrule

\multicolumn{1}{l|}{MAM \cite{Chen_MAM_arxiv2020}} & 
\multicolumn{1}{l}{2020} & 
\multicolumn{1}{|l|}{Speech} & 
\multicolumn{1}{l|}{Spectrogram} & 
\multicolumn{1}{l}{ST} \\

\midrule

\multicolumn{1}{l|}{FAT-MLM \cite{Zheng_FATMLM_ICML2021}} & 
\multicolumn{1}{l}{2021} & 
\multicolumn{1}{|l|}{Speech, $<$Text-Speech$>$} & 
\multicolumn{1}{l|}{Spectrogram} & 
\multicolumn{1}{l}{ASR, ST, MT} \\

\midrule

\multicolumn{1}{l|}{A$^3$T \cite{Bai_A3T_ICML2022}} & 
\multicolumn{1}{l}{2022} & 
\multicolumn{1}{|l|}{Speech, $<$Text-Speech$>$} & 
\multicolumn{1}{l|}{Spectrogram} & 
\multicolumn{1}{l}{TTS, Speech Editing} \\

\bottomrule
\label{SpeechPretraining}  
\end{tabular}}
\end{table}

\subsection{Adversarial Training}
Adversarial training is a technique to improve the robustness of deep learning models by exposing them to adversarial examples, which are modified inputs designed to cause incorrect predictions. One common adversarial training approach is through a GAN \cite{GoodFellow_GAN_NIPS2014}, where a generator and a discriminator are trained simultaneously. The generative model creates outputs that follow the distribution of the input dataset, while the discriminator tries to discern between the generated sample and the actual input dataset. TTS systems employ adversarial training to improve specific aspects, such as the speech's naturalness or the speaker similarity of the generated speech.

\subsection{Meta Learning}
Meta-learning, or learning to learn, is a strategy to speed up the learning process from a small amount of data. Model-Agnostic Meta-Learning (MAML) \cite{Finn_ICML_2017} explores this strategy across different models and problem settings, such as few-shot image classification and few-shot regression. For voice cloning, It identifies an effective parameter initialization for accelerating speaker adaptation. It utilizes an outer loop to determine a meta-initialization and an inner loop to quickly adapt to a new task using minimal data. Language-Agnostic Meta-Learning (LAML) treats languages as independent tasks while building multilingual voice cloning systems \cite{Huang_MetaTTS_TASLP2022}.

\section{Voice Cloning}
\label{Section4}

Voice cloning is a complex problem, as improving its performance requires tackling different aspects within TTS systems, such as speech naturalness, quality, expressiveness, and speaker similarity. Typically, these modifications occur in acoustic modeling and, in some instances, in the vocoder. We will explore the contributions, following Fig. \ref{VoiceCloning_SurveyOverview}, as understanding these methodologies will pave the way toward enhancing voice cloning and understanding their limitations.

\subsection{Speaker Adaptation}

\begin{figure}
\centering
  \includegraphics[width=1.00\linewidth]{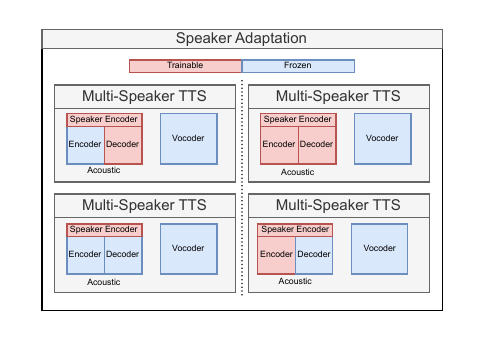}
\caption{Four variants for speaker adaptation of a Multi-Speaker TTS.}
\label{VoiceCloning_finetuning_process}  
\end{figure}

Speaker Adaptation relies on fine-tuning a TTS model to replicate the voice of a certain speaker. This process is one of the main principles behind voice cloning. The fine-tuning process, summarized in Fig. \ref{VoiceCloning_finetuning_process}, can happen on the speaker encoder, a model component, or the entire model.

\textbf{Disentanglement:} Representation disentanglement is an active area of research exploring decomposing the TTS system's elements to achieve better control of voice characteristics. Daft-Exprt \cite{Zaidi_DaftExprt_InterSpeech2022} introduced an acoustic model for cross-speaker prosody transfer based on FastSpeech 2. This model uses FiLM conditioning layers \cite{Perez_Film_AAAI2018} to inject prosodic information across the model, encoding high-level and low-level prosodic information (pitch, loudness, duration). This model includes an adversarial training strategy through a speaker classifier for speaker-prosody disentanglement. Ellinas et al. \cite{Ellinas_SpeechCom2023} proposed a phoneme-level prosody control for TTS systems using Tacotron as its base acoustic model. They introduce discrete labels, achieved through an unsupervised prosodic clustering process, to discretize prosodic features such as the fundamental frequency (f0) and phoneme duration. These features will be fed into a prosody encoder to obtain prosodic representation. Additionally, they use an adversarial speaker classifier for phoneme-speaker disentanglement. Du et al. \cite{Du_TASLP2023} presented TN-VQTTS, a novel TTS model that leverages timbre-normalized vector-quantized (TN-VQ) acoustic features for style-timbre disentanglement in speaker adaptation with limited data. This architecture consists of two main components: a txt2vec module that predicts TN-VQ features from input phonemes and a vec2wav module that uses those features, auxiliary prosodic features, and a speaker embedding from a conditioned speaker encoder to generate a speaker-specific waveform. TN-VQ features are derived by performing dimensionality reduction of the VQ features via PCA, followed by timbre normalization. The resulting features are then quantized using K-means clustering for integration into a multi-speaker TTS model. For speaker adaptation, they use one of two approaches: fine-tuning free adaptation, where they use a pre-trained style embedding lookup table (LUT) as a codebook and choose the most suitable embedding, or weight fine-tuning adaptation, where it performs a weighted summation of all the embeddings within the table. 

\textbf{Speaker Representation and Verification:} Speaker adaptation has been evolving over the years. Thus, researchers explore speaker representation and speaker-related tasks such as speaker verification to improve speaker similarity. Fan et al. \cite{Fan_ICASSP2015} proposed a multi-speaker DNN that shares parameters across various speakers with a speaker-dependent regression layer. A new speaker is adapted by fixing the shared parameters and updating the regression layer. Zhao et al. \cite{Zhao_InterSpeech2016} proposed a unified framework for speaker adaptation that includes a bidirectional Long Short-Term memory with a Recurrent Neural Network (Bi-LSTM-RNN) acoustic model. They explored speaker identity representations, such as i-vector and speaker code, at the input layer of the framework to observe the effectiveness of different speaker representations. Doddipatla et al. \cite{Doddipatla_InterSpeech2017} explored the usage of speaker identification vectors, called d-vectors instead of the previously used i-vectors, with linguistic features to train a multi-speaker DNN-based TTS model. More recent work started exploring the usage of speaker verification to enhance speaker adaptation performance. Cai et al. \cite{Cai_InterSpeech2020} proposed a method for improved knowledge transfer from a speaker verification network to a TTS framework, using a feedback constraint mechanism to facilitate voice cloning training. Their approach integrates a Tacotron-based model and a learnable dictionary encoding (LDE)-based verification model \cite{Cai_LDE_TASLP2020} to enhance speaker similarity for unseen speakers. Ruggiero et al. \cite{Ruggiero_arxiv2021} proposed a multi-speaker TTS system consisting of three-component: a speaker encoder, a synthesizer, and a vocoder, following the general architecture of Jia et al. \cite{Jia_NIPS2018}. They explore various architectures for the speaker encoder and a novel transfer-learning technique that conditions the synthesizer on utterance embeddings instead of speaker embeddings. This approach enables the system to generalize to unseen speakers using only a short reference utterance. Nakai et al. \cite{Nakai_APSIPAASC2022} extends on GANSpeech \cite{Yang_GANSpeech_InterSpeech2021} and FastSpeech 2 by introducing a novel multi-task adversarial training algorithm through discerning real from fake synthesized speech and performing speaker verification to ensure whether the speaker exists or is non-existent. Additionally, they introduce Adversarially Constrained Autoencoder Interpolation (ACAI), a regularization term to improve performance for unseen speakers. Chen et al. \cite{Chen_TASLP_NFVC_2022} tackled voice cloning by proposing a neural fusion architecture to include unit concatenation within a parametric TTS. This fusion architecture comprises a text encoder, acoustic decoder, and phoneme-level reference encoder. This phoneme-level encoder extracts the speakers' prosody to improve speaker similarity and uses auto-regressive distribution modeling and decoder refinement to tackle concatenation discontinuity. Wadoux et al. \cite{Wadoux_TSD2023} explored the impact of phonetic content and sample duration on voice cloning. The system consists of a multi-speaker Tacotron 2 model and an x-vector speaker encoder. This exploration led to inconclusive results since the impact is shown in the speaker representation but not the generated waveform; these results showed improved speech quality and speaker similarity in very specific cases.

\textbf{Handling Speech Variability:} Speaker adaptation can face difficulties handling speech variability from different speaking styles, accents, and acoustic conditions. GMVAE-Tacotron \cite{Hsu_ICLR2019} focused on speech variabilities, such as controlling speech attributes such as speaking style, accent, noise, and acoustic conditions. It introduced the VAE-based hierarchical generative TTS model, improving attribute control compared to the Global Style token (GST)-based approaches \cite{Wang_GST_ICML2018, Akuzawa_InterSpeech2018}. They explored synthesizing speech for unseen speakers in clean and noisy environments from an unseen reference utterance. Similarly, Neekhara et al. \cite{Neekhara_arxiv2021} tackled the challenge of noisy speech in multi-speaker public datasets by proposing a transfer-learning guideline for adapting a single-speaker TTS model for a new speaker. More specifically, two fine-tuning methods are used to reduce the time and data required to adapt the TTS model to a new voice. AdaSpeech \cite{Adaspeech_ICLR2021} introduced a novel architecture, based on FastSpeech 2, for efficient and high-quality speaker adaptation in TTS systems. It models acoustic information at both the utterance and phoneme levels to handle acoustic variety and enable better generalization across various acoustic conditions. Additionally, it employs conditional layer normalization (CLN) in the mel-spectrogram decoder, improving the efficiency during speaker adaptation by only fine-tuning the CLN parameters and speaker embeddings during speaker adaptation. AdaSpeech 3 \cite{AdaSpeech3_InterSpeech2021} builds upon AdaSpeech but focuses on synthesizing spontaneous speech, similar to podcasts and conversation, instead of reading-style speech. It utilizes the already available reading-style TTS and fine-tunes it for spontaneous TTS. Adapting to spontaneous TTS includes inserting filled pauses (FP) through an FP predictor, a mixture of experts (MoE)-based duration predictor, and a pitch predictor for rhythm adaptation. Finally, the authors fine-tune specific decoder parameters for target timbre speaker adaptation. Mandeel et al. \cite{Mandeel_SpeD2023} concentrated on improving prosody control, especially in interrogative sentences. This work explores speaker adaptation using FastSpeech 2 with a HiFi-GAN neural vocoder, in which they demonstrated their performance on two HiFi-based adaptation datasets \cite{Bakhturina_HiFiTTS_InterSpeech2021} containing interrogative and declarative sentences. 

\textbf{Untranscribed Speech:} The abundance of untranscribed speech online can be a useful tool for improving voice cloning systems. Inoue et al. \cite{Inoue_ICASSP2020} proposed a semi-supervised speaker adaptation method utilizing a pre-trained ASR system to obtain text data from available untranscribed speech recordings and then use it to train a TTS model. This method can simplify speaker adaptation through an end-to-end ASR/TTS system. Zhang et al. \cite{Zhang_InterSpeech2020} introduced a novel unsupervised pre-training mechanism that uses VQ-VAE \cite{VanDenOord_VQVAE_NIPS2017} to extract linguistic units directly from untranscribed speech and then use $<$linguistic unit, speech$>$ pairs to pre-train Tacotron. For speaker adaptation, we use a few $<$text, audio$>$ pairs for fine-tuning to target speakers. AdaSpeech 2 \cite{Adaspeech2_ICASSP2021} is an adaptive TTS that uses untranscribed speech for speaker adaptation. It follows the basic structure of AdaSpeech and utilizes a mel-spectrogram encoder for speech reconstruction, constrained by an L2 loss with the phoneme encoder. For speaker adaptation, untranscribed speech is used to fine-tune the TTS decoder. Klapsas et al. \cite{Klapsas_InterSpeech2022} explored BYOL \cite{Grill_BYOL_NIPS2020}, a self-supervised framework for obtaining speech representation. Then, they extended the augmentations typically applied through training to capture the speakers' identity and ensure the features are noise-resistant. These pre-trained weights are used as utterance-level embeddings and inputs to a Non-Attentive Tacotron-based architecture \cite{Shen_NonAttentiveTacotron_arxiv2020} to be used for unseen speaker adaptation.

\textbf{Parameter Efficiency:} Personalized TTS focuses on reducing TTS models' complexity to ensure usability in less computational devices. Inoue et al. \cite{Inoue_APSIPA2020} investigated the roles of various modules in Transformer-TTS \cite{Li_TransformerTTS_AAAI2019} during fine-tuning to understand its effectiveness in handling speech deletion and/or repetition. This includes freezing specific modules, including character embedding, encoder, probability output layer, and choosing to include Binary Cross Entropy (BCE) loss in the fine-tuning process. The authors concluded that fine-tuning the TTS model while freezing character embeddings and the probability output layer reduces speech repetition and deletion. AdaVITS \cite{Song_AdaVITS_ISCSLP_2022} is a lightweight VITS-based TTS for low-resource speaker adaptation. They replace the upsampling-based decoder with an Inverse short-time Fourier transform (iSTFT)-based decoder. They also introduce NanoFlow to reduce the prior encoder parameters and replace scaled-dot attention with linear attention to reduce the text encoder complexity. Phonetic posteriorgram (PPG) is used as the frame-level linguistic feature for supervision as the model becomes less stable. HyperTTS \cite{Li_HyperTTS_LREC_COLING2024} proposed a hypernetwork-based approach for TTS speaker adaptation, integrating a small learnable network into an adapted FastSpeech 2 backbone. Instead of fine-tuning the entire model, the hypernetwork dynamically generates adapter parameters conditioned on speaker embeddings derived from a speaker verification model trained with generalized end-to-end loss (GE2E) \cite{Wan_GE2E_ICASSP2018}.

\textbf{Non-English TTS:} Voice cloning works mainly focus on English speaker adaptation, even though there are many languages worldwide. VStyclone \cite{Wu_VStyclone_CEE2023}, a real-time Mandarin voice cloning algorithm, consists of three main components: an efficient tone extractor, a transformer-based style synthesizer with an embedded style extraction module, and a GAN-based vocoder. The tone extractor uses self-attention to capture speaker-specific information efficiently instead of GE2E loss optimized with a modified softmax loss. Cheng et al. \cite{Cheng_ICDSP2022} proposed using prosodic features for Mandarin speaker adaptation. These features are obtained at the utterance/speaker level and then integrated into existing speaker-encoding or speaker-embedding adaptation frameworks such as Tacotron 2.

\begin{figure*}
\centering
  \includegraphics[width=0.96\linewidth]{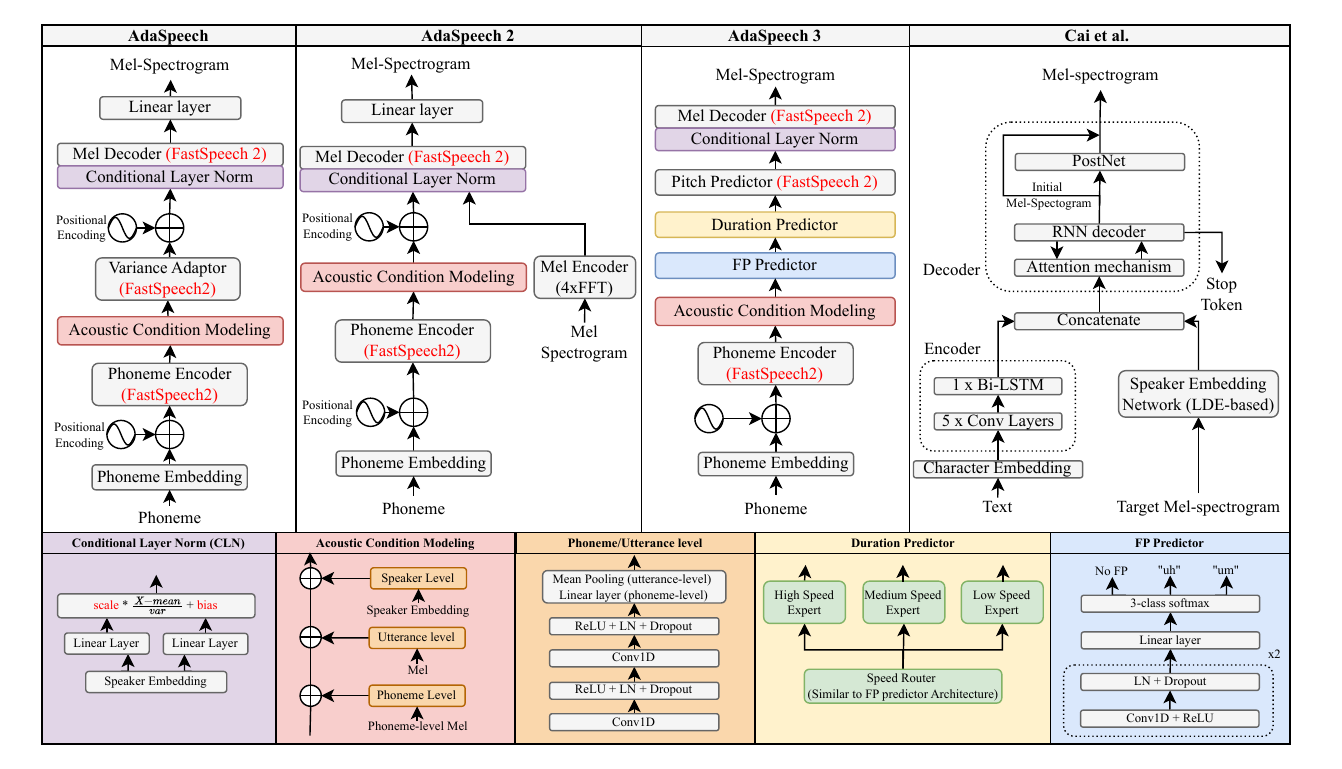}
\caption{Few-Shot TTS algorithms, adapted from AdaSpeech \cite{Adaspeech_ICLR2021}, AdaSpeech 2 \cite{Adaspeech2_ICASSP2021}, AdaSpeech 3 \cite{AdaSpeech3_InterSpeech2021}, and Cai et al. \cite{Cai_InterSpeech2020}.}
\label{CommonFewShotTTS}  
\end{figure*}

\subsection{Few-shot Voice Cloning}

Few-Shot TTS (FS-TTS), namely Few-Shot Voice Cloning, focuses on speaker adaptation through fine-tuning with minimal data, which is crucial in tackling practical voice cloning cases. Fig. \ref{CommonFewShotTTS} shows some few-shot TTS algorithms.

\textbf{Disentanglement:} Researchers actively study methods for obtaining better control of speech elements, including content-speaker disentanglement, to improve few-shot learning performance. Wang et al. \cite{Wang_InterSpeech2020} proposed a few-shot speaker adaptation method through spoken content and voice factorization to tackle overfitting due to the complexity of acoustic models. They decompose the acoustic model into a spoken content predictor that predicts PPG to constrain the acoustic model space. Then, we use it as an input to a voice conversion module to extract the acoustic features. Additionally, they proposed speaker adaptation strategies for data with or without transcription. Attentron \cite{Choi_Attentron_InterSpeech2020} is a Tacotron2-based novel architecture for few-shot voice cloning, where it utilizes two encoders to improve unseen speaker similarity and speech quality. This architecture includes a fine-grained encoder that extracts style information from multiple reference audio files, and the coarse-grained encoder generates global embeddings of the target speaker, which helps stabilize the output. It follows a warm-start training method, similar to \cite{Cooper_ICASSP2020}, to reduce training time. Lu et al. \cite{Lu_ICASSP2024} proposed an end-to-end unified model for voice cloning and conversion as they share a similar information flow. The authors proposed extending VAE-based speech disentanglement \cite{Lu_SLT2022, Lian_ICASSP2022} by introducing learnable text-aware prior, enhancing content-speaker disentanglement by providing more accurate and speaker-independent content representation than the initially used fixed Gaussian prior.

\textbf{Untranscribed Speech:} Few-shot TTS systems require text-speech pairs to perform fine-tuning to clone a speaker's voice. Researchers propose voice cloning algorithms without requiring transcribed speech. Nautilus \cite{Luong_Nautilus_TASLP2020} is a multi-speaker TTS/VC system with the ability of speaker adaptation using untranscribed speech through a learned linguistic latent embedding (LLE). The process of voice cloning inference without transcribed speech follows these steps: they first adapt the speech decoder and vocoder to the target's voice speech. Then, both are welded together by joint fine-tuning and, finally, inference. Similarly, Sadekova et al. \cite{Sadekova_InterSpeech2022} built a multimodal TTS/VC system capable of performing voice cloning and any-to-any voice conversion using untranscribed data. This system consists of three modules: a text encoder, a mel decoder, and a diffusion-based shared decoder. Zhang et al. \cite{Zhang_ICASSP2022} also proposed a multimodal learning approach capable of both TTS and VC. They extend Tacotron 2 by introducing an unsupervised speech representation module, VQ-VAE, to extract linguistic units given speech input. Then, a Tacotron 2-based model generates a mel spectrogram conditioned on these unsupervised units for VC or phonemes for TTS to generate a waveform through a vocoder. On the contrary, other works focus on uni-modal TTS algorithms, HierSpeech \cite{Lee_HierSpeech_NIPS2022} builds on top of VITS by employing self-supervised learning to reduce the text-speech information gap. To improve linguistic feature extraction, they utilize XLS-R \cite{Babu_XLSR_InterSpeech2022}, a pre-trained Wav2Vec 2.0 on large-scale cross-lingual speech dataset. They further extend this architecture by introducing HierSpeech-U, which performs speaker adaptation with untranscribed speech. Guided-TTS 2 \cite{Kim_GuidedTTS2_arxiv2022} builds on top of Guided-TTS \cite{Kim_GuidedTTS_ICML2022} and enables adaptive TTS using untranscribed data. Guided-TTS proposed a diffusion-based TTS model that uses an unconditional Denoising Diffusion Probabilistic Model (DDPM) \cite{Ho_DDPM_NIPS2020}, phoneme classifier, duration predictor, and a speaker encoder to remove the reliance on text-speech alignment by using untranscribed data. Additionally, it introduces a norm-based scaling method to improve pronunciation. Then, Guided-TTS 2 extends this work by introducing a speaker-conditional diffusion model to enable few-shot adaptation and zero-shot TTS for target speakers. This model fine-tunes the pre-trained speaker conditional diffusion model. Due to the difficulty of training its unconditional generative model, Kim et al. \cite{Kim_UnitSpeech_InterSpeech2023} developed a method to fine-tune an adaptive diffusion-based TTS model, UnitSpeech, with limited untranscribed data. It utilizes a unit encoder that generates a discretized representation from untranscribed speech using HuBERT \cite{Hsu_Hubert_TASLP2021}. The output representation is then used for speaker adaptation, improving speech quality and speaker similarity.

\textbf{Parameter Efficiency:} Personalized TTS involves researchers' contributions toward reducing models' sizes and improving their speed. For example, Voiceloop \cite{Taigman_Voiceloop_ICLR2018} proposed a unique approach that deals with samples in the wild without requiring phoneme alignment or linguistic features. This work is based on a novel shifting buffer working memory that estimates the attention, computes output, and updates the buffer. This architecture can fit new speakers post-training by only updating the speaker embeddings, and it can be run almost in real-time on a CPU. Similarly, Kons et al. \cite{Kons_InterSpeech2019} proposed a lightweight TTS model that consists of three modules: prosody prediction, acoustic feature prediction, and an LPCNet vocoder. This system can run faster than real-time on a standard CPU. Morioka et al. \cite{Morioka_arxiv2022} proposed a parameter-efficient few-shot speaker adaptation method. It uses a PnG NAT model \cite{Jia_PnGBERT_InterSpeech2021, Shen_NonAttentiveTacotron_arxiv2020} as its backbone model and then includes trainable lightweight modules called residual adapters for each target speaker while keeping the backbone architecture frozen. Similarly, Hsieh et al. \cite{Hsieh_InterSpeech2023} proposed parameter-efficient adapter modules, using FastPitch \cite{Lancucki_Fastpitch_ICASSP2021} as a baseline TTS model through only fine-tuning those adapters to a target speaker during speaker adaptation. Huang et al. \cite{Huang_ICASSP2023} proposed trainable structured pruning for voice cloning, utilizing FastSpeech 2 as its backbone, to compress the voice cloning model up to 7 times while maintaining its performance. Chen et al. \cite{Chen_DiT_TTS_SSW2023} introduced diffusion-based backbones for adaptive TTS called Diffusion Transformer (DiT). They utilize adaptive layer normalization (adaLN), initially introduced for image generation \cite{Peebles_DiT_CVPR2023}, and made it compatible by receiving a sequence as a condition instead of class embeddings. In addition, they use a generator-based diffusion method to enable faster and higher-quality speech synthesis, demonstrating superior performance compared to transformer-based solutions. Recently, Wang et al. \cite{Wang_USAT_TASLP2024} proposed a universal speaker-adaptive TTS (USAT) that unifies few-shot and zero-shot strategies. The framework introduces a timbre converter consisting of two discriminators and a Memory-Augmented VAE (MA-VAE) to generate a timbre-invariant phoneme representation to enable disentanglement during pre-training. USAT demonstrates high parameter efficiency as it requires adapting fewer parameters while maintaining FS-TTS performance.

\textbf{Speaker Representation:} Improving speaker representation extraction is crucial to enhance the overall performance of TTS systems, specifically speaker similarity, for unseen speaker adaptation. Deng et al. \cite{Deng_Arxiv2018} explored the multi-speaker latent space to enhance few-shot speaker adaptation using just a few minutes of data or improving voice quality by leveraging data from multiple speakers. GC-TTS \cite{Kim_GCTTS_SMC2021} proposed a Tacotron 2 extension by introducing a neural speaker encoder and a classification layer. The authors utilized two geometric constraints, initially proposed in \cite{Jung_TNNLS2020} for natural images, to improve speaker representation. This system employs a multi-stage training and fine-tuning approach while exploring freezing-specific modules for a more efficient fine-tuning process. Lee et al. \cite{Lee_InterSpeech2022} proposed a novel few-shot adaptation algorithm that utilizes a speaker embedding LUT for training a multi-speaker Tacotron 2 TTS system. They introduce a GAN-based initial embedding predictor trained on speaker embeddings using adversarial training. This approach speeds up the adaptation process by generating the initial embeddings suitable for unseen speaker adaptation. Voice Filter \cite{Gabrys_VoiceFilter_ICASSP2022} takes another approach by shifting the voice speaker adaptation task into two modules: a high-quality TTS system followed by a voice conversion system as a post-processing step. This split leads to requiring less amount of data for speaker adaptation by limiting the problem's complexity. Xue et al. \cite{Xue_ECAPA_TDNN_ISCSLP2022} proposed a three-module separately trained TTS architecture consisting of an ECAPA-TDNN speaker encoder \cite{ECAPA_TDNN_InterSpeech2020}, FastSpeech 2 acoustic model, and a HiFi-GAN vocoder. Similarly, TDNN-VITS \cite{Zhao_TDNN_VITS_ICSP2023} adopts the same speaker encoder for the VITS model. It introduces a multi-angle feature fusion approach, which improves speech quality with less data rather than simply concatenating speaker embeddings after its text embeddings.

\textbf{Optimization-based:} Improving FS-TTS performance is achievable through developing novel architectures or even components to enhance its performance. Additionally, exploring optimization techniques such as meta-learning and Bayesian optimization is an exciting approach to enhance performance with minor architectural modifications. Chen et al. \cite{Chen_ICLR2018} introduced a meta-learning approach applied to a multi-speaker model with a shared conditional WaveNet core. Their training mechanism focuses on having two sets of parameters: task-dependent and independent. During training, they train all but discard the dependent ones for deployment; this leads to only optimizing for the task-dependent parameters for faster adaptation to new tasks. Moss et al. \cite{Moss_BOFFINTTS_ICASSP2020} proposed Bayesian Optimization for Fine Tuning Neural Text to Speech (BOFFIN TTS), in which achieving convincing results requires an adaptation strategy. This strategy includes fine-tuning adaptation hyper-parameters and adding two new ones, considering the target speaker's acoustic, phonetic, and corpus properties to achieve optimal results. Min et al. \cite{Min_Meta_StyleSpeech_ICML2021} proposed StyleSpeech, a FastSpeech2-based model capable of synthesizing speech in the style of a target speaker using a single reference speech sample. The authors introduce style-adaptive layer normalization (SALN), which aligns the gain and bias of the input text features based on the style vector. Building on this, they introduce Meta-StyleSpeech by incorporating style prototypes and episodic meta-learning to enhance the model's performance to unseen speakers. Unlike previous works, Meta-Voice \cite{Liu_MetaVoice_arxiv2021} explores MAML and meta-transfer for faster few-shot speaker adaptation. This architecture, based on FastSpeech 2, consists of four modules: a text encoder, a variance adapter, a mel decoder, and a style encoder. They also use and extend SALN to handle multiple speakers and prosodies. In addition, it uses domain adversarial training for speaker-prosody disentanglement, which is useful for cross-speaker style transfer. Similarly, Meta-TTS \cite{Huang_MetaTTS_TASLP2022} proposed applying MAML to speed up few-shot speaker adaptation using FastSpeech 2 as its base TTS architecture. The authors achieve this by designing a meta-task, where each task randomly chooses a speaker and takes ten utterances from that speaker, 5 for the support set and 5 for the query. This training consists of an inner loop and an outer loop, where the inner loop optimizes specific modules using multiple gradient descents (inner loop update), and the outer loop uses second-order gradient computation for the loss (Meta update gradient).

\begin{table*}[t]
\centering
\caption{Comparison of Few-shot TTS algorithms covering naturalness (Nat), Speech quality (Qual), and speaker similarity (Sim). For multiple datasets, results are averaged. Papers are arranged in ascending order by publication date.}
\resizebox{18cm}{!}{
\begin{tabular}{l|l|l}

\toprule

\multicolumn{1}{c|}{\textbf{Paper}} & \multicolumn{1}{c|}{\textbf{Test set, additional details}} & \multicolumn{1}{c}{\textbf{TTS Evaluation}} \\ 

\midrule

\rowcolor{lightgray}
\multicolumn{3}{c}{Naturalness and Speech Quality} \\

\midrule

\multirow{2}{*}{Attentron \cite{Choi_Attentron_InterSpeech2020}} 
&
\multirow{2}{*}{VCTK (10 utt/spk, 8 spks)} 
&
\underline{\resizebox{1.5cm}{!}{Nat (NMOS) $\uparrow$}}: Attentron (3.97) $>$ Cooper et al. \cite{Cooper_ICASSP2020} (3.91) $>$ GMVAE-Tacotron (3.88) \\ 
& & 
\underline{\resizebox{1.5cm}{!}{Qual (MCD) $\downarrow$}}: Attentron (11.67) $>$ GMVAE-Tacotron (13.11) $>$ Cooper et al. (13.50) \\

\midrule

\multirow{3}{*}{Meta-StyleSpeech (MSS) \cite{Min_Meta_StyleSpeech_ICML2021}} 
&
\multirow{3}{*}{VCTK (1 utt/spk, 108 spks)} 
&
\underline{\resizebox{1.5cm}{!}{Nat (NMOS) $\uparrow$}}: MSS (3.82) $>$ SS (3.77) $>$ MS-FS2 (d-vector) (3.74) $>$ MS-FS2 (3.69) $>$ GMVAE \cite{Hsu_ICLR2019} (3.15)\\ 
& & 
\underline{\resizebox{1.5cm}{!}{Qual (MCD) $\downarrow$}}: MSS (4.95) $>$ MS-FS2 (4.97) $>$ SS (5.01) $>$ MS-FS2 (d-vector) (5.03) $>$ GMVAE (5.54)\\
& & 
\underline{\resizebox{1.5cm}{!}{Qual (WER) $\downarrow$}}: MSS (16.79) $>$ MS-FS2 (17.35) $>$ SS (17.51) $>$ MS-FS2 (d-vector) (17.55) $>$ GMVAE (23.86) \\

\midrule

\multirow{2}{*}{GC-TTS \cite{Kim_GCTTS_SMC2021}} 
&
\multirow{1}{*}{VCTK (80 utt, 16 spks) [Subjective]; VCTK (800 utt, 16 spks) [Objective]} 
&
\multirow{1}{*}{\underline{\resizebox{1.5cm}{!}{Nat (NMOS) $\uparrow$}}: Cooper et al. \cite{Cooper_ICASSP2020} (3.86) $>$ Tacotron 2 (3.84) $>$ Jia et al. \cite{Jia_NIPS2018} (3.83) $>$ GC-TTS* (3.81) $>$ Tacotron 2* (3.74)} \\ 
& 
\multirow{1}{*}{*5 minutes fine-tuning}
&
\multirow{1}{*}{\underline{\resizebox{1.5cm}{!}{Qual (MCD) $\downarrow$}}: Tacotron 2 (4.87) $>$ GC-TTS* (5.35) $>$ Tacotron 2* (5.38) $>$ Jia et al. (5.77) $>$ Cooper et al. (5.82)}\\

\midrule

\multirow{1}{*}{Meta-TTS \cite{Huang_MetaTTS_TASLP2022}} & 
LibriTTS (1 utt/spk, 30 speakers); VCTK (1 utt/spk, 80 spks)&

\multirow{1}{*}{\underline{\resizebox{1.5cm}{!}{Nat (NMOS) $\uparrow$}}: Meta-TTS (3.20) $>$ MS-FS2 (3.16) [emb table] | MS-FS2 (3.72) $>$ Meta-TTS (2.70) [Shared emb]}\\ 

\midrule

\multirow{1}{*}{Sadekova et al. \cite{Sadekova_InterSpeech2022}} 
&
\multirow{1}{*}{VCTK (5 utt/spk, 25 spks)}
&
\underline{\resizebox{1.5cm}{!}{Nat (NMOS) $\uparrow$}}: Tacotron-SMA \cite{Popov_InterSpeech2020} (4.23) $>$ Proposed (4.18) $>$ FS (3.98) $>$ Grad-TTS (3.94) $>$ SS (3.77)\\ 

\midrule

\multirow{2}{*}{UnitSpeech \cite{Kim_UnitSpeech_InterSpeech2023}} 
&
\multirow{2}{*}{LibriTTS test-clean (5 utt/spk, 10 spks)}
&
\underline{\resizebox{1.5cm}{!}{Nat (NMOS) $\uparrow$}}: GTTS2 \cite{Kim_GuidedTTS2_arxiv2022} (4.16) $>$ UnitSpeech (4.13) $>$ YTTS (3.57)\\ 
& & 
\underline{\resizebox{1.5cm}{!}{Qual (CER) $\downarrow$}}: GTTS2 (0.84) $>$ UnitSpeech (1.75) $>$ YTTS (2.38) \\

\midrule

\multirow{1}{*}{TDNN-VITS \cite{Zhao_TDNN_VITS_ICSP2023}} 
&
\multirow{1}{*}{VCTK (300 utt, 10 spk)}
&
\underline{\resizebox{1.5cm}{!}{Nat (NMOS) $\uparrow$}}: TDNN-VITS (3.72) $>$ FS2 (3.61) $>$ Tacotron 2 (3.54) \\ 

\midrule

\multirow{2}{*}{VAE-TP \cite{Lu_ICASSP2024}} 
&
\multirow{2}{*}{LibriTTS test set (50 utt/spk, 11 spks)}
 &
\underline{\resizebox{1.5cm}{!}{Nat (NMOS) $\uparrow$}}: CDFSE (3.82) $>$ VAE-TP (3.81) $>$ SC-GlowTTS (3.56) \\ 
& & 
\underline{\resizebox{1.5cm}{!}{Qual (CER) $\downarrow$}}: CDFSE (1.96) $>$ VAE-TP (2.82) $>$ SC-GlowTTS (6.99) \\

\midrule

\multirow{2}{*}{USAT \cite{Wang_USAT_TASLP2024}} 
&
\multirow{2}{*}{ESLTTS (10 utt/spk, 30 speakers) | *VITS is fully fine-tuned}
&
\underline{\resizebox{1.5cm}{!}{Nat (NMOS) $\uparrow$}}: USAT (3.84) $=$ UnitSpeech (3.84) $>$ VITS (3.80)\\ 
& &
\underline{\resizebox{1.5cm}{!}{Qual (WER) $\downarrow$}}: UnitSpeech (10.30) $>$ USAT (11.00) $>$ VITS (14.60)\\

\midrule

\rowcolor{lightgray}
\multicolumn{3}{c}{Speaker Similarity} \\

\midrule

\multirow{2}{*}{Attentron \cite{Choi_Attentron_InterSpeech2020}} 
&
\multirow{2}{*}{VCTK (10 utt/spk, 8 spks)} 
&
\underline{\resizebox{1.5cm}{!}{Sim (SMOS) $\uparrow$}}: Attentron (3.57) $>$ GMVAE (3.27) $>$ Cooper et al. (3.17) \\
& & 
\underline{\resizebox{1.5cm}{!}{Sim (SECS) $\uparrow$}}: Attentron (0.79) $>$ Cooper et al. (0.71) $>$ GMVAE (0.70) \\

\midrule

\multirow{2}{*}{Meta-StyleSpeech (MSS) \cite{Min_Meta_StyleSpeech_ICML2021}} 
&
\multirow{2}{*}{VCTK (1 utt/spk, 108 spks)} 
&
\underline{\resizebox{1.5cm}{!}{Sim (SMOS) $\uparrow$}}: MSS (3.81) $>$ SS (3.46) $>$ MS-FS2 (3.36) $>$ GMVAE (3.11) $>$ MS-FS2 (d-vector) (2.12)\\
& & 
\underline{\resizebox{1.5cm}{!}{Sim (SECS) $\uparrow$}}: MSS (0.82) $>$ SS (0.80) $>$ MS-FS2 (0.77) $=$ GMVAE (0.77) $>$ MS-FS2 (d-vector) (0.62)\\

\midrule

\multirow{2}{*}{GC-TTS \cite{Kim_GCTTS_SMC2021}} 
&
\multirow{1}{*}{VCTK (80 utt, 16 spks) [Subjective]; VCTK (800 utt, 16 spks) [Objective]} 
&
\multirow{1}{*}{\underline{\resizebox{1.5cm}{!}{Sim (SMOS) $\uparrow$}}: Tacotron 2 (3.60) $>$ GC-TTS* (3.45) $>$ Tacotron 2* (3.34) $>$ Cooper et al. (3.19) $>$ Jia et al. (3.15)}  \\
& 
\multirow{1}{*}{*5 minutes fine-tuning}
&
\multirow{1}{*}{\underline{\resizebox{1.5cm}{!}{Sim (SECS) $\uparrow$}}: Tacotron 2 (0.89) $>$ GC-TTS* (0.84) $>$ Tacotron 2* (0.82) $>$ Cooper et al. (0.80) $>$ Jia et al. (0.78)}  \\

\midrule

\multirow{1}{*}{Meta-TTS \cite{Huang_MetaTTS_TASLP2022}} & 
LibriTTS (1 utt/spk, 30 speakers); VCTK (1 utt/spk, 80 spks)&

\multirow{1}{*}{\underline{\resizebox{1.5cm}{!}{Sim (SMOS) $\uparrow$}}: Meta-TTS (2.96) $>$ MS-FS2 (1.55) [emb table] | Meta-TTS (3.06) $>$ MS-FS2 (1.33) [Shared emb]}\\

\midrule

\multirow{1}{*}{Sadekova et al. \cite{Sadekova_InterSpeech2022}} 
&
\multirow{1}{*}{VCTK (5 utt/spk, 25 spks)}
&
\underline{\resizebox{1.5cm}{!}{Sim (SMOS) $\uparrow$}}: Tacotron-SMA (4.31) $>$ Grad-TTS (4.19) $>$ Proposed (4.15) $>$ FS (4.02) $>$ SS (3.96)\\

\midrule

\multirow{2}{*}{UnitSpeech \cite{Kim_UnitSpeech_InterSpeech2023}} 
&
\multirow{2}{*}{LibriTTS test-clean (5 utt/spk, 10 spks)}
&
\underline{\resizebox{1.5cm}{!}{Sim (SMOS) $\uparrow$}}: GTTS2 (3.90) $=$ UnitSpeech (3.90) $>$ YTTS (3.34)\\
& & 
\underline{\resizebox{1.5cm}{!}{Sim (SECS) $\uparrow$}}: GTTS2 (0.94) $=$ UnitSpeech (0.94) $>$ YTTS (0.87)\\

\midrule

\multirow{1}{*}{TDNN-VITS \cite{Zhao_TDNN_VITS_ICSP2023}} 
&
\multirow{1}{*}{VCTK (300 utt, 10 spk)}
&
\underline{\resizebox{1.5cm}{!}{Sim (SMOS) $\uparrow$}}: TDNN-VITS (3.85) $>$ FS2 (3.73) $>$ Tacotron 2 (3.65) \\

\midrule

\multirow{2}{*}{VAE-TP \cite{Lu_ICASSP2024}} 
&
\multirow{2}{*}{LibriTTS test set (50 utt/spk, 11 spks)}
 &
\underline{\resizebox{1.5cm}{!}{Sim (SMOS) $\uparrow$}}: VAE-TP (3.66) $>$ CDFSE (3.60) $>$ SC-GlowTTS (3.38)\\
& & 
\underline{\resizebox{1.5cm}{!}{Sim (SECS) $\uparrow$}}: VAE-TP (0.78) $>$ CDFSE (0.76) $>$ SC-GlowTTS (0.72)\\

\midrule

\multirow{2}{*}{USAT \cite{Wang_USAT_TASLP2024}} 
&
\multirow{2}{*}{ESLTTS (10 utt/spk, 30 speakers) | *VITS is fully fine-tuned}
&
\underline{\resizebox{1.5cm}{!}{Sim (SMOS) $\uparrow$}}: UnitSpeech (3.75) $>$ USAT (3.74) $>$ VITS (3.56)\\
& &
\underline{\resizebox{1.5cm}{!}{Sim (SECS) $\uparrow$}}: UnitSpeech (0.83) $=$ USAT (0.83) $>$ VITS (0.80)\\

\bottomrule

\end{tabular}}
\label{FewShotTTS_comparison}
\end{table*}

\textbf{Non-English FS-TTS:} Mandarin is the second language with voice cloning publications. Cong et al. \cite{Cong_InterSpeech2020} performed speech-noise disentanglement to improve performance with noisy audio target samples by introducing domain adversarial training. Zheng et al. \cite{Zheng_ICASSP2021} used a Tacotron2-based architecture and proposed various extensions, such as a novel speaker encoder with a random sampling mechanism. They also used a speaker classifier, and text encoder adversarial training for content-speaker disentanglement. Finally, they also proposed multi-alignment guided attention for text-audio pairs for model training stability. Wang et al. \cite{Wang_ICASSP2021} proposed a speaker adaptation framework that disentangles prosody and voice characteristics. They achieved this by introducing a prosody control attention mechanism and an LPCNet vocoder to ensure computational efficiency. In addition, they proposed a pruning strategy to remove mismatching text-audio pairs, ensuring a more stable system. Fu et al. \cite{Fu_ICASSP2021} proposed separate speaker and prosody modeling based on a reference audio and PPG using a multi-head attention process, allowing more control of prosody duration and speaker timbre. In addition, the decoder is factored into an average-net decoder to control prosody by generating speaker-independent acoustic features and an adaptation-net decoder to generate acoustic features conditioned on speaker embeddings for speaker timbre imitation. Dian \cite{Song_Dian_ICASSP2021} is a speech synthesis system consisting of an attention-free acoustic model with a duration model. The attention was replaced by inputting the duration directly into the acoustic model, eliminating the skipping and repeating challenge and improving speech intelligibility. Chien et al. \cite{Chien_ICASSP2021} investigated various speaker representations, including pre-trained and learnable ones. This study concluded that speaker embeddings pre-trained through voice conversion achieved the highest performance compared to d/x-vectors. Additionally, learnable speaker representations, such as speaker embedding LUT and GST, can improve speaker similarity. He et al. \cite{Ke_AdaptiveFormer_ISCSLP2022} proposed AdaptiveFormer, a FastSpeech2-based FS-TTS model, with two contributions: proposing a novel adaptive attention mechanism that employs speaker embeddings, modified calculations for Q, K and V, and CLN compared of the traditional multi-head self-attention component. They also introduce a conditional adapter to predict rhythm information (duration, pitch, energy). UNet-TTS \cite{Li_UNetTTS_ICASSP2022} is a one-shot TTS algorithm replicating unseen speakers' voices and styles. It adopts a UNet-based architecture, initially proposed for image segmentation \cite{Ronneberger_UNet_MICCAI2015}, to efficiently extract hierarchical speaker and utterance level features from a reference audio. In addition, the authors introduced a content loss after the style encoder to enhance content-style disentanglement. Unlike Tacotron baselines using GST or speaker embedding obtained from a speaker verification task, this method enables fine-grained prediction of acoustic features and better performance on an unseen emotional corpus. MSDTRON \cite{Wu_Msdtron_ICASSP2022} proposed a multi-speaker TTS system to tackle speaking style diversity. This paper investigated effective approaches toward making use of characteristic information. This includes explicitly characterizing the harmonic structure of speech through an excitation spectrogram and a conditional gated LSTM (CGLSTM) to improve voice control without affecting speech quality. LightClone \cite{Wu_LightClone_InterSpeech2023} focused on speeding up speaker adaptation and inference time by introducing a speaker-guided subnet selection process. They use FastSpeech 2 as the baseline model, and during adaptation, only one subnet of the source model is selected for the target speaker. Unlike previous work, AdaVocoder  \cite{Yuan_AdaVocoder_InterSpeech2022} approached voice cloning by proposing a custom voice vocoder with a cross-domain consistency loss to tackle overfitting in few-shot voice cloning.

\textbf{Summary:} Few-shot TTS algorithms have been rapidly evolving, where research focuses on novel components, architectures, or training strategies to enhance the overall performance. Table \ref{FewShotTTS_comparison} shows a performance comparison of few-shot TTS algorithms in naturalness, speech quality, and speaker similarity. Word Error Rate (WER) and Character Error Rate (CER) are objective metrics useful to ascertain speech quality by testing it using a pre-trained Automatic Speech Recognition (ASR) system. This system can vary, such as using Whisper ASR \cite{Radford_Whisper_ICML2023}, CTC-based conformer \cite{Gulati_Conformer_Interspeech2020} available in Nemo toolkit \cite{Kuchaiev_Nemo_arxiv2019}. The same goes for Speaker Embeddings Cosine Similarity (SECS), as the speaker encoder used for reporting results varies. This includes X-vector \cite{Snyder_X_Vector_ICASSP2018}, GE2E \cite{Wan_GE2E_ICASSP2018} obtained through the Resemblyzer package, and TitaNet-L \cite{Koluguri_Titanet_ICASSP2022}.

\begin{figure*}
\centering
  \includegraphics[width=0.96\linewidth]{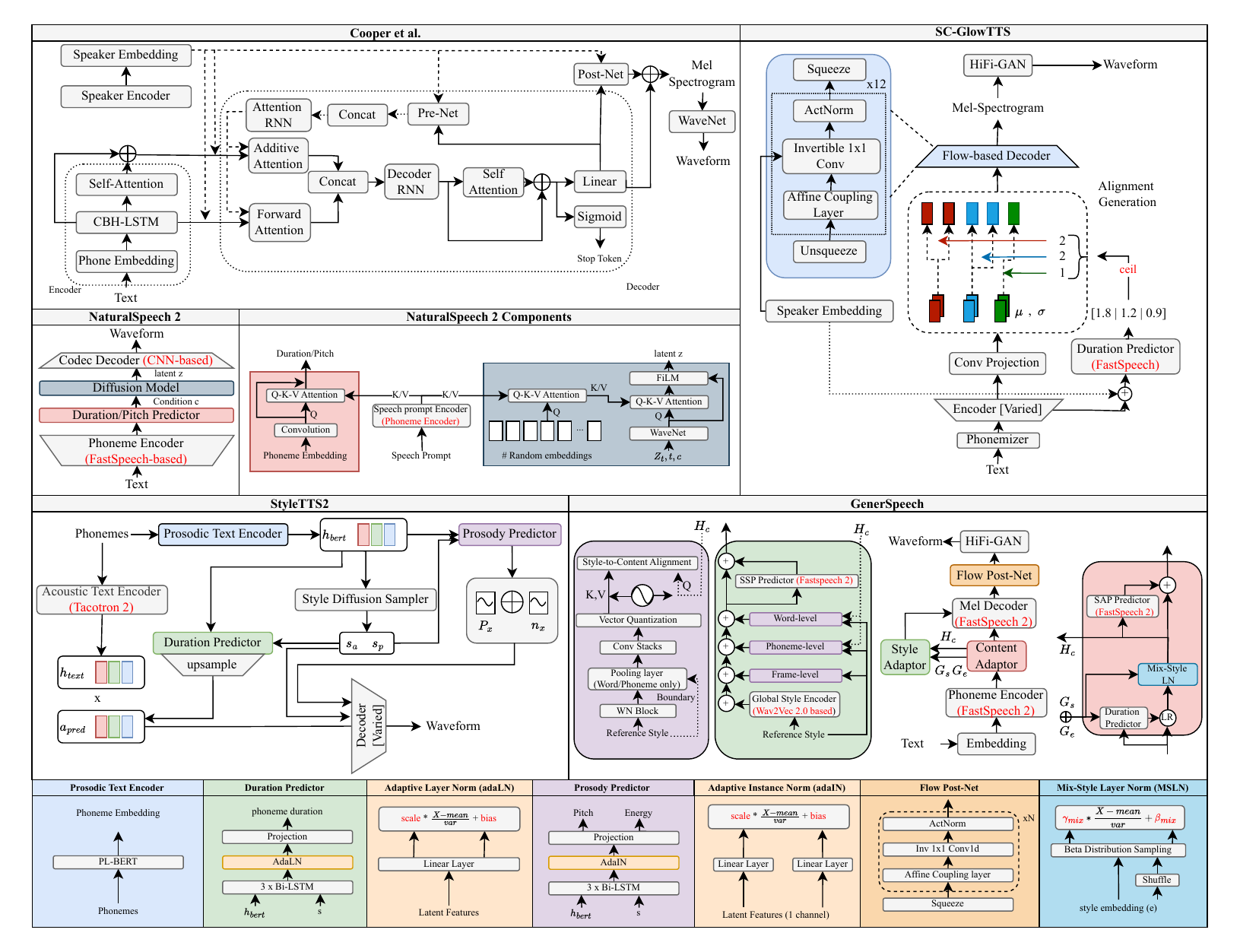}
\caption{Zero-shot TTS algorithms, adapted from SC-GlowTTS \cite{Casanova_SCGlowTTS_InterSpeech2021}, StyleTTS 2 \cite{Li_StyleTTS2_NIPS2024}, NaturalSpeech 2 \cite{Shen_NaturalSpeech2_ICLR2024}, GenerSpeech \cite{Huang_Generspeech_NIPS2022}, and Cooper et al. \cite{Cooper_ICASSP2020}}
\label{ZeroShotTTS_Figure1}  
\end{figure*}

\subsection{Zero-shot Voice Cloning}

Zero-shot TTS (ZS-TTS) algorithms aim to generate speech that closely resembles a target speaker without fine-tuning the TTS model. Unlike speaker adaptation, ZS-TTS architectures require the inclusion of specialized modules, such as a speaker encoder, to capture and replicate the target speaker's voice during inference. Fig. \ref{ZeroShotTTS_Figure1} illustrates some ZS-TTS algorithms.

\textbf{Disentanglement:} Representation disentanglement is an exciting approach as it provides more control over the target voice's attributes. GenerSpeech \cite{Huang_Generspeech_NIPS2022} decomposes style-agnostic (content) from style-specific (speaker identity, emotion, prosody) information while papers \cite{Choi_SPL2024, Jeon_AAAI2024, Chen_MRMITTS_ACM_ALRLIP2024, Wang_GZS_TV_InterSpeech2023, Lian_UTTS_TASLP2023} focus on timbre-content disentanglement. GenerSpeech is a FastSpeech2-based TTS model that tackles disentanglement by introducing multiple modules, such as a multi-level style adapter with a Wav2Vec 2.0 encoder. In addition, it uses a mix-style layer normalization (MSLN) to eliminate style from content representation. GZS-TV \cite{Wang_GZS_TV_InterSpeech2023} proposed a generalizable zero-shot TTS and VC model that uses a disentangled representation learning mechanism to separate phoneme from timbre embeddings to reduce phoneme leakage into speaker embeddings. The authors built an ECAPA-TDNN-based speaker encoder \cite{ECAPA_TDNN_InterSpeech2020} and leveraged VAE to enhance its performance. UTTS \cite{Lian_UTTS_TASLP2023} is an unsupervised TTS acoustic model training scheme that does not require text-audio pairs. They employ a conditional disentangled sequential VAE (C-DSVAE) as the model's backbone for timbre-content disentanglement. Jeon et al. \cite{Jeon_AAAI2024} proposed a FastSpeech-based zero-shot TTS algorithm that utilizes a negation feature learning paradigm that decouples speaker information from linguistic content by subtracting content representations from the audio embedding. This approach reduces content leakage into speaker representation, leading to more accurate speaker identity representation. They also introduce multi-stream transformers to improve robustness to unseen speakers. Similarly, MRMI-TTS \cite{Chen_MRMITTS_ACM_ALRLIP2024} is a Fastspeech2-based framework for zero-shot TTS that reduces content leakage in speaker embeddings using a multi-reference audio approach. It employs a speaker-content disentangling module (SCDM) comprising a separate content and speaker encoder. The multi-reference audio selection is based on Jaccard similarity to ensure that the content is varied and that mutual information minimizes the entanglement between speaker and content embeddings. Choi et al. \cite{Choi_SPL2024} proposed a novel speaker conditioning technique using a variable length embedding sequence and enhancing an affine coupling function for flow-based ZS-TTS architectures. This method builds upon the VITS model as its base architecture. They tackle timbre-content disentanglement by using different references and target speech from the speaker. Unlike previous methods, Ren et al. \cite{Ren_ICASSP2024} proposed TiCodec, a neural speech codec that enhances efficiency by separating time-variant and invariant speech features. By encoding time-invariant features separately, it reduces the number of tokens needed while maintaining performance in ZS-TTS.

Other researchers take a different approach by emphasizing finer control over speech synthesis. Paul et al. \cite{Paul_InterSpeech2021} proposed a multi-speaker and multi-style TTS framework capable of generating speech with the speaker's characteristics given a reference signal. They proposed a novel Renyi Divergence-based disentangled representation framework to disentangle content, style, and speaker information, ensuring reduced content and style leakage. Kumar et al. \cite{Kumar_TASLP2022} proposed a novel zero-shot TTS approach that uses a normalization architecture with a NAR transformer-based architecture. They proposed two normalization techniques to improve learning prosodic properties, such as energy and f0, in a disentangled manner. Fujita et al. \cite{Fujita_ICASSPW2023} proposed a zero-shot TTS model conditioned on speech embeddings derived from a self-supervised learning (SSL) model. They also introduced the separate conditioning of acoustic features and the phoneme duration predictor, effectively disentangling rhythm and acoustic features. Lee et al. \cite{Lee_InterSpeech2023} disentangled timbre (intra-speaker consistent component) from cadence (inter-utterance variate component) through a hierarchical structure to model each independently, and then used variance-invariance covariance regularization (VICReg) on cadence embeddings and a speaker ID loss to ensure stable timbre embeddings for the speaker. HierSpeech++ \cite{Lee_HierSpeechPlusPlus_arxiv2023} is a fully parallel zero-shot TTS and VC framework built on top of HierSpeech \cite{Lee_HierSpeech_NIPS2022}. They proposed multiple improvements, such as the Text-to-Vec (TTV) module that generates semantic and f0 representation from text and prosody prompts. It also introduced an efficient speech super-resolution (SpeechSR) component for efficient up-sampling of high-resolution audio waveforms. HierSpeech++ enables fine-grained control over speech synthesis by disentangling speaker-related and speaker-agnostic representation, including an encoder with f0 control for manual control of pitch contour. Mega-TTS \cite{MegaTTS_arxiv2023} is a novel large-scale zero-shot TTS system that enhances speech generation by considering intrinsic speech properties. It achieves this by disentangling speech into four components: content, timbre, prosody, and phase, using mel spectrogram as the intermediate representation instead of codec-based features. The model employs a timbre encoder as a global representation since timbre changes slowly over time, while a VQGAN-based acoustic model generates a spectrogram and a prosody large language model (P-LLM) to capture prosody distribution, effectively modeling both local and long-range dependencies. Mega-TTS 2 \cite{jiang_MegaTTS2_ICLR2024} builds on top of its previous work by proposing a multi-sentence prompting method to extract diverse speaker information. The authors designed a compressive acoustic autoencoder for timbre-prosody disentanglement, achieving more control while maintaining the speaker's identity. They also introduced a multi-reference timbre encoder (MRTE) and a P-LLM for fine-grained timbre information extraction from multi-reference prompts.

Codec-based TTS systems use a hierarchical approach to obtain semantic tokens followed by acoustic tokens. Even though they are not fully independent, they aim to decompose speech into semantic (linguistic and para-linguistic) and acoustic (non-linguistic, e.g., speaker identity) information. Papers \cite{SPEAR_TTS_ACL2023, Zhu_VecTok_Speech_arxiv2023, Xue_InterSpeech2024, Lee_InterSpeech2024, Ju_NaturalSpeech3_ICML2024, Lei_ICASSP2024} are codec-based TTS algorithms that disentangle semantic and acoustic features. Kharitonov et al. \cite{SPEAR_TTS_ACL2023} introduced SPEAR-TTS, a multi-speaker TTS system that formulates speech synthesis as two seq2seq tasks: converting text into high-level semantic tokens (Reading) and then mapping semantic tokens into low-level acoustic tokens (Speaking). This decomposition enables training the reading stage with text-audio pairs while the speaking module leverages self-supervised audio models obtained from speech-only data. Vec-Tok Speech \cite{Zhu_VecTok_Speech_arxiv2023} is a multi-task generation framework capable of zero-shot VC, TTS, and Speech-to-Speech Translation (S2ST). It introduced a novel speech codec with an encoder-decoder architecture, where the encoder extracts speech vectors and semantic tokens. Moreover, an autoregressive language model predicts semantic tokens, followed by byte-pair encoding (BPE), to reduce exposure bias and improve context coverage. The Semantic tokens remove speaker-related information while preserving its content. Similarly, Xue et al. \cite{Xue_InterSpeech2024} proposed a novel codec-based TTS algorithm that improves context utilization and handles longer speech prompts. They introduced multi-modal context-enhanced Qformer (MMCE-Qformer), building on Q-former \cite{Li_Blip2_ICML2023}, to utilize multi-modal context information. They also leveraged a pre-trained LLM to predict semantic tokens and use the SoundStorm model \cite{Borsos_SoundStorm_arxiv2023} to predict acoustic tokens, thereby improving speech quality and speaker similarity. Lee et al. \cite{Lee_InterSpeech2024} introduced a novel two-stage TTS framework. The text goes through an interpreting module based on Token Transducer++, a modified version of Token Transducer \cite{Kim_TokenTransducer_arxiv2024}, to obtain semantic tokens and a speaking module based on a conformer-based architecture with a Grouped Masked Language Model (G-MLM) \cite{Jeong_GMLM_SPL2024}, to convert those tokens into acoustic ones. Finally, a codec decoder converts them into a speech waveform. Lei et al. \cite{Lei_ICASSP2024} proposed a novel TTS algorithm using multi-scale acoustic prompts to better capture the timbre and speaking style of the target speaker. Additionally, they use a speaker-aware text encoder and a modified VALL-E acoustic decoder to improve speaker similarity while maintaining high-quality speech. NaturalSpeech 3 \cite{Ju_NaturalSpeech3_ICML2024} is a novel diffusion-based method that factorizes speech into various attributes (content, prosody, timbre, acoustic details) by introducing a novel factorized diffusion model and a neural codec with factorized vector quantization (FVQ). These contributions improved the speech's naturalness, intelligibility, prosody, and speaker similarity.

\textbf{Emotion Transfer:} Maintaining the emotional characteristics of the reference audio and, in some instances, even controlling the emotional state of the synthesized speech is crucial toward expressive speech generation. StyleTTS \cite{Li_StyleTTS_arxiv2022} is a style-based generative modeling TTS algorithm that synthesizes speech with the prosody and emotion of the reference audio. It introduced a transferable Monotonic Aligner (TMA) for better text-speech alignment and duration-invariant data augmentation for improved duration prediction. In addition, it leveraged adaptive instance normalization (AdaIN) to incorporate style vectors, enabling natural and diverse speech generation. StyleTTS 2 \cite{Li_StyleTTS2_NIPS2024} builds upon StyleTTS by proposing the usage of style diffusion models, adversarial training, and a pre-trained WavLM \cite{Chen_WavLM_JSTSP2022} during training, and a pre-trained PL-BERT \cite{Li_PLBERT_ICASSP2023} to enable natural and expressive speech. It performs like VALL-E on zero-shot speaker adaptation but uses 250 times less data, and can synthesize emotion-specific speech without a reference speech. DINO-VITS \cite{Pankov_DINO_VITS_InterSpeech2024} introduced a data-efficient ZS-TTS system that utilizes a self-supervised loss to enhance robustness in noisy conditions. It captures intra-speaker variation in speaking style and emotion, proven through an emotion recognition classifier on speaker embeddings. This architecture consists of an mBART-based Text-to-Unit (T2U) module for feature extraction \cite{Liu_mBART_ACL2020}, a speaker encoder based on CAM++ \cite{Wang_CAMPlusPlus_InterSpeech2023} speaker verification model and a VITS-based Unit-to-Speech (U2S) module. DINO-VITS employs a multi-objective training approach integrating self-supervised DINO loss alongside speech synthesis objective losses to improve speaker representation learning and noise robustness. On the other hand, ZET-Speech \cite{Kang_ZET_Speech_InterSpeech2023} differs from previous models as it is an emotion-controllable TTS model requiring a neutral reference speech and an emotional label to generate emotion-specific audio. Its architecture is based on Grad-StyleSpeech and uses adversarial training and guidance methods for emotion-speaker disentanglement.

\textbf{Parameter Efficiency:} Ensuring the parameter efficiency of voice cloning algorithms can assist in easier deployment without requiring specialized hardware. Paul et al. \cite{Paul_SC_WaveRNN_InterSpeech2020} proposed a speaker-conditional WaveRNN (SC-WaveRNN) universal vocoder that efficiently performs speech generation of unseen speakers and acoustic conditions. It is a WaveRNN-variant that uses explicit speaker information, such as speaker embeddings, to improve performance without speaker adaptation or retraining the network, demonstrating enhanced speech naturalness and speaker similarity in zero-shot TTS. SC-GlowTTS \cite{Casanova_SCGlowTTS_InterSpeech2021} builds on GlowTTS \cite{Kim_GlowTTS_NIPS2020} to achieve an efficient zero-shot multi-speaker TTS model. They use an external speaker encoder based on Angular prototypical loss \cite{Chung_InterSpeech2020}, an adapted HiFi-GAN v2 vocoder, and explored the performance while using Glow TTS's transformer-based encoder, residual dilated convolutional network \cite{Vainer_SpeedySpeech_InterSpeech2020} and a gated convolutional network \cite{Dauphon_ICML2017}. Similarly, Ogun et al. \cite{Ogun_InterSpeech2023} build on Glow-TTS to improve utterance diversity by introducing an explicit stochastic pitch predictor and a flow-based duration predictor. Yoon et al. \cite{Yoon_InterSpeech2023} introduced sparse attention, a transformer pruning method, to improve the TTS algorithm's generalizability. They selected StyleSpeech as the baseline model and then implemented sparse attention on the decoder only as it has more self-attention layers compared to the encoder, and applying it to the encoder degrades the performance. This cuts the redundant self-attention layer connections with weights below an automatically set threshold. Lux et al. \cite{Lux_SLT2023} used FastSpeech 2 as the base TTS model and proposed an utterance-level normalization and utterance-level speaker embeddings. They also introduced a lightweight aligner for fine-grain prosodic feature extraction and fine-tuning within seconds. TacoLM \cite{Song_TacoLM_InterSpeech2024} is a two-stage codec-based TTS model consisting of EnCodec \cite{Defossez_encodec_TML2023} as an audio tokenizer and a BPE encoder to extract text representation. The authors improved inference efficiency and reduced model size by introducing two contributions: a gated attention mechanism as a replacement for regular multi-head attention used in transformer-based language models and a gated cross-attention layer to improve computational and storage efficiency. CM-TTS \cite{CM_TTS_NAACL2024} is a generative novel architecture based on consistency models to reduce the steps required to achieve high-quality speech synthesis. The authors also introduced weighted samplers to ensure unbiased learning through the training process. Similarly, FlashSpeech \cite{Ye_FlashSpeech_ACMMM2024} is a novel adversarial consistency training approach to train a latent consistency model from scratch without a pre-trained diffusion model. Additionally, it introduced a prosody generator to improve the diversity of the speech synthesized. This model achieves high-quality performance 20 times faster than other TTS systems. Fujita et al. \cite{Fujita_InterSpeech2024} proposed a lightweight system using a mixture of adapters (MoA) to enhance speaker adaptation while maintaining high-quality speech synthesis with faster inference. These modules are integrated into the decoder and the variance adapter of a non-autoregressive TTS model such as FastSpeech 2. SimpleSpeech \cite{Yang_SimpleSpeech_InterSpeech2024} is a NAR diffusion-based TTS system trained with speech-only data without alignment information and can generate speech directly from text. It proposed a speech-codec model based on scalar quantization that maps speech into a compact latent space, reducing complexity and increasing generation speed. Small-E \cite{Lemerle_SmallE_InterSpeech2024} is a codec-based language model that relies on an encoder-decoder architecture in contrast to a decoder-only mechanism. It introduced Linear Casual Language Model (LCLM) blocks as a replacement for AR transformers, leading to efficient training. The authors also proposed position-aware cross-attention (PACA) tackling skipping \& repeating errors. P-Flow \cite{Kim_PFlow_NIPS2024} is a ZS-TTS model that consists of a speech-prompted text encoder within a NAR TTS model and a flow-based generative decoder for efficient speech synthesis. This text encoder enables maintaining the emotional prosody of the reference speech for synthesis.

\textbf{Generalizable Speech Synthesis:} The versatility of Speech synthesis models is crucial to handling various tasks such as voice conversion, speech editing, and zero-shot TTS, or the diverse speaking styles. Glow-WaveGAN 2 \cite{GlowWaveGAN2_InterSpeech2022} expands on Glow-WaveGAN \cite{GlowWaveGAN_InterSpeech2021} to enable zero-shot TTS and any-to-any VC. This architecture includes a Universal WaveGAN as the feature extractor and the vocoder, a multi-speaker GlowTTS, and an external speaker encoder. UnifySpeech \cite{Liu_UnifySpeech_arxiv2023} combines TTS and VC tasks into one framework by categorizing speech into three independent components: content, speaker, and prosody information. Everything in the model is shared except for the speech content extraction. Nansy++ \cite{Choi_NANSY_ICLR2023} is a unified framework for various speech tasks such as VC, TTS, SVS, and voice design. Its backbone consists of an analyzer and synthesizer trained with speech-only data. Nansy++ analyzer consists of separate modules for pitch, linguistic, and timbre information that aim to disentangle the linguistic features from all speech representations. Ke et al. \cite{Ke_Voicifier_ICAIPR2023} proposed a FastSpeech2-based novel TTS method, Voicifier, to improve zero-shot TTS performance across multiple accents. Voicifier uses high-frequency mel spectrogram bins and two strategies to maintain timbre features while discarding content and rhythm. Then, they perform voice adaptation through Voice-Pin layer normalization (VPLN) to integrate timbre features with text features for better speaker adaptation. Zhang et al. \cite{Zhang_SPL2023} proposed a zero-shot multi-accent TTS model. They introduce an accent classifier on the Tacotron 2 framework to have more control of pronunciation variety, which is applicable for generating accented speech for unseen speakers. SpeechX \cite{SpeechX_TASLP2024} is a versatile speech generation model capable of zero-shot TTS for clean and noisy speech samples. They leverage NCLM by extending VALL-E with multi-task learning through task-dependent prompts. VoiceCraft \cite{VoiceCraft_ACL2024} is a token-infilling NCLM capable of speech editing and ZS-TTS. It leveraged a transformer-decoder architecture and introduced a token rearrangement process to enable inter-sequence generation. Kanda et al. \cite{Kanda_elate_arxiv2024} proposed an Expressive Laughter-controllable ZS-TTS Engine (ELaTE) to integrate laughing sounds, built on top voicebox \cite{Le_Voicebox_NIPS2024}, a conditional flow-matching-based ZS-TTS, and condition it using a laugh detector through fine-tuning.

\begin{table*}
\centering
\caption{Comparison of Zero-shot TTS algorithms covering naturalness (Nat) and Speech Quality (Qual). For multiple datasets, results are averaged. Papers are arranged in ascending order by publication date.}
\resizebox{18cm}{!}{
\begin{tabular}{l|l|l}

\toprule

\multicolumn{1}{c|}{\textbf{Paper}} & \multicolumn{1}{c|}{\textbf{Test set, additional details}} & \multicolumn{1}{c}{\textbf{TTS Evaluation}} \\ 

\midrule

\multirow{1}{*}{DiffVoice \cite{Liu_DiffVoice_ICASSP2023}} 
& 
\multirow{1}{*}{LibriTTS (5 utt/reference, 21 reference audios) | [*P=Prompt]} 
&
\multirow{1}{*}{\underline{\resizebox{1.5cm}{!}{Nat (NMOS) $\uparrow$}}: DiffVoice-P (4.66) $>$ VITS (x-vector) (4.14) $>$ YTTS (3.46) $>$ FS2 (x-vector) (3.42) $>$ MSS (2.82)} \\

\midrule

\multirow{2}{*}{GZS-TV \cite{Wang_GZS_TV_InterSpeech2023}} 
&
\multirow{2}{*}{LibriTTS test-set (5 utt/spk, 37 spks); VCTK (2 utt/spk, 108 spks)} 
&
\underline{\resizebox{1.5cm}{!}{Nat (NMOS) $\uparrow$}}: GZS-TV (3.76) $>$ YTTS (3.65) $>$ SS (3.22) $>$ MSS (3.20) \\ 
& &
\underline{\resizebox{1.5cm}{!}{Qual (WER) $\downarrow$}}: GZS-TV (8.55) $>$ YTTS (8.95) $>$ SS (9.65) $>$ MSS (9.85) \\

\midrule

\multirow{1}{*}{STEN-TTS \cite{Tran_STEN_TTS_InterSpeech2023}} 
&
\multirow{1}{*}{VCTK (110 spks)} 
&
\underline{\resizebox{1.5cm}{!}{Nat (NMOS) $\uparrow$}}: STEN-TTS (3.01) $>$ DiffSinger \cite{Liu_DiffSinger_AAAI2022} (2.55) $>$ SS (2.50)\\

\midrule

\multirow{2}{*}{HierSpeech++ (HS++) \cite{Lee_HierSpeechPlusPlus_arxiv2023}} 
&
\multirow{1}{*}{LibriTTS test-clean (4837 samples)} 
& 
\underline{\resizebox{1.5cm}{!}{Nat (NMOS) $\uparrow$}}: HS++ (4.55) $>$ HS (4.05) $>$ VALL-E X (3.52) $>$ YTTS (3.51) $>$ XTTS (3.17) \\ 
& 
\multirow{1}{*}{LibriTTS test-other (5120 samples)} 
& 
\underline{\resizebox{1.5cm}{!}{Qual (WER) $\downarrow$}}: HS++ (5.20) $>$ HS (7.88) $>$ YTTS (11.97) $>$ XTTS (28.89) $>$ VALL-E X (32.88)\\

\midrule

\multirow{2}{*}{Latent Filling \cite{Bae_LatentFilling_ICASSP2024}} 
&
\multirow{2}{*}{VCTK (11 spks)}  & 
\underline{\resizebox{1.5cm}{!}{Nat (NMOS) $\uparrow$}}: Proposed (3.82) $>$ YTTS (3.16) $>$ SC-GlowTTS (2.02) \\ 
& &
\underline{\resizebox{1.5cm}{!}{Qual (WER) $\downarrow$}}: Proposed (1.10) $>$ YTTS (4.78) $>$ SC-GlowTTS (28.56) \\

\midrule

\multirow{2}{*}{Mega-TTS 2 (MTTS2) \cite{jiang_MegaTTS2_ICLR2024}} 
& 
\multirow{2}{*}{LibriSpeech test-clean (100 sec/spk, 20 spks)} 
&
\underline{\resizebox{1.5cm}{!}{Qual (QMOS) $\uparrow$}}: MTTS2-300s (4.12) $>$ MTTS2-3s (3.99) $>$ VALL-E-3s (3.89) $>$ VALL-E-20s (3.41) \\ 
& & 
\underline{\resizebox{1.5cm}{!}{Qual (WER) $\downarrow$}}: MTTS2-300s (2.23) $>$ MTTS2-3s (2.46) $>$ VALL-E-3s (5.83) $>$ VALL-E-20s (8.77) \\

\midrule

\multirow{2}{*}{USAT \cite{Wang_USAT_TASLP2024}} 
&
\multirow{2}{*}{ESLTTS, VCTK, LibriTTS test set (10 utt/spk, 30 spks) each} 
&
\multirow{1}{*}{\underline{\resizebox{1.5cm}{!}{Nat (NMOS) $\uparrow$}}: USAT (3.89) $>$ YTTS (3.73) $>$ SS (3.32) $>$ MSS (3.30)}\\ 
& & 
\multirow{1}{*}{\underline{\resizebox{1.5cm}{!}{Qual (WER) $\downarrow$}}: USAT (10.40) $>$ YTTS (10.70) $>$ SS (11.20) $>$ MSS (11.70)} \\

\midrule

\multirow{2}{*}{MRMI-TTS \cite{Chen_MRMITTS_ACM_ALRLIP2024}} 
&
\multirow{2}{*}{VCTK (9 spks); LibriTTS test-clean} 
&
\underline{\resizebox{1.5cm}{!}{Nat (NMOS) $\uparrow$}}: MRMI-TTS (3.97) $>$ MSS (3.73) $>$ SS (3.70) $=$ FS2 (d-vector) (3.70) \\ 
& &
\underline{\resizebox{1.5cm}{!}{Qual (MCD) $\downarrow$}}: MRMI-TTS (5.02) $>$ MSS (5.17) $>$ SS (5.28) $=$ FS2 (d-vector) (5.28) \\

\midrule

\multirow{1}{*}{P-Flow \cite{Kim_PFlow_NIPS2024}} 
&
\multirow{1}{*}{LibriSpeech test-clean (4-10 sec/sample/spk, 40 spks) = 2.2h [similar to VALL-E]} 
&
\multirow{1}{*}{\underline{\resizebox{1.5cm}{!}{Qual (WER) $\downarrow$}}: P-Flow (2.60) $>$ VALL-E (5.90) $>$ YTTS (7.70)}\\

\midrule

\multirow{1}{*}{StyleTTS 2 (STTS2) \cite{Li_StyleTTS2_NIPS2024}} 
& 
\multirow{1}{*}{LibriTTS test-clean (500 samples)} 
&
\underline{\resizebox{1.5cm}{!}{Nat (NMOS) $\uparrow$}}: STTS2 (4.15) $>$ STTS+HiFi-GAN (3.91) $>$ VITS (3.69) $>$ YTTS (2.35) \\ 

\midrule

\multirow{1}{*}{VALL-E 2 \cite{VALLE2_arxiv2024}} 
&
\multirow{1}{*}{LibriSpeech test-clean (4-10 sec/sample/spk, 40 spks) = 2.2h | *GroupSize = 1} 
& 
\multirow{1}{*}{\underline{\resizebox{1.5cm}{!}{Qual (WER) $\downarrow$}}: VALL-E 2 [FTS] (0.65) $>$ VALL-E [FTS] (0.75) $>$ VALL-E 2 (1.50) $>$ VALL-E (3.10)} \\

\midrule

\multirow{3}{*}{CM-TTS \cite{CM_TTS_NAACL2024}} 
&
\multirow{3}{*}{VCTK (512 samples); LJSpeech (512 samples)} 
&
\underline{\resizebox{1.5cm}{!}{Qual (QMOS) $\uparrow$}}: CM-TTS (T=4) (3.82) $>$ DiffGAN-TTS \cite{Liu_DiffGAN_arxiv2022} (T=4) (3.66) \\ 
& &
\underline{\resizebox{1.5cm}{!}{Qual (MCD) $\downarrow$}}: DiffGAN-TTS \cite{Liu_DiffGAN_arxiv2022} (T=4) (8.66) $>$ CM-TTS (T=4) (8.90) \\
& &
\underline{\resizebox{1.5cm}{!}{Qual (WER) $\downarrow$}}: DiffGAN-TTS \cite{Liu_DiffGAN_arxiv2022} (T=4) (7.98) $>$ CM-TTS (T=4) (8.97) \\

\midrule

\multirow{2}{*}{NaturalSpeech 3 (NS3) \cite{Ju_NaturalSpeech3_ICML2024}} 
& 
\multirow{1}{*}{LibriSpeech test-clean (1 utt/spk, 40 spks)} 
&
\underline{\resizebox{1.5cm}{!}{Nat (CMOS) $\uparrow$}}: NS3 (reference) $>$ NS2 (-0.18) $>$ MTTS2 (-0.20) $>$ VB (-0.32) $>$ VALL-E (-0.60)\\ 
& 
\multirow{1}{*}{*All are trained using LibriLight dataset} 
& 
\underline{\resizebox{1.5cm}{!}{Qual (WER) $\downarrow$}}: NS3 (1.81) $>$ NS2 (1.94) $>$ VB (2.14) $>$ MTTS2 (2.32) $>$ VALL-E (6.11) \\ 

\midrule

\multirow{1}{*}{SpeechX \cite{SpeechX_TASLP2024}} 
&
\multirow{1}{*}{LibriSpeech test-clean (4-10 sec/sample/spk, 40 spks) = 2.2h [similar to VALL-E]} 
&
\multirow{1}{*}{\underline{\resizebox{1.5cm}{!}{Qual (WER) $\downarrow$}}: SpeechX (VALL-E init) (4.66) $>$ SpeechX (Random init) (5.40) $>$ VALL-E (5.90)}\\

\midrule

\multirow{2}{*}{VoiceCraft \cite{VoiceCraft_ACL2024}} 
& 
\multirow{1}{*}{LibriTTS (125 utt); YouTube-GigaSpeech test-set (125 utt)} 
&
\multirow{1}{*}{\underline{\resizebox{1.5cm}{!}{Nat (NMOS) $\uparrow$}}: VoiceCraft (4.17) $>$ XTTS v2 \cite{Eren_Coqui2021} (3.96) $>$ VALL-E (3.86) $>$ YTTS (2.79)}\\ 
& 
\multirow{1}{*}{Objective|Subjective: 125|40 utt/dataset}
&
\multirow{1}{*}{\underline{\resizebox{1.5cm}{!}{Qual (WER) $\downarrow$}}: XTTS v2 \cite{Eren_Coqui2021} (3.60) $>$ VoiceCraft (4.50) $>$ YTTS (6.60) $>$ VALL-E (7.10)}  \\
\midrule

\multirow{2}{*}{Xue et al. \cite{Xue_InterSpeech2024}} 
& 
\multirow{1}{*}{LibriSpeech test-clean (1249 sentences)} 
&
\underline{\resizebox{1.5cm}{!}{Nat (NMOS) $\uparrow$}}: Proposed (3.79) $>$ XTTS (3.44) $>$ VALL-E (3.24)\\
& 
\multirow{1}{*}{IEMOCAP (1079 sentences)} 
&
\underline{\resizebox{1.5cm}{!}{Qual (MCD) $\downarrow$}}: Proposed (7.72) $>$ XTTS (7.95) $>$ VALL-E (8.76) \\

\midrule

\multirow{2}{*}{Lee et al. \cite{Lee_InterSpeech2024}} 
&
\multirow{2}{*}{LibriTTS test-clean (500 sentences)} 
&
\underline{\resizebox{1.5cm}{!}{Nat (NMOS) $\uparrow$}}: Proposed (3.94) $>$ VALL-E X (3.75) $>$ VITS (ECAPA TDNN) (3.52)\\ 
& &
\underline{\resizebox{1.5cm}{!}{Qual (CER) $\downarrow$}}: Proposed (2.34) $>$ VITS (ECAPA TDNN) (4.81) $>$ VALL-E X (8.72)\\

\midrule

\multirow{3}{*}{SimpleSpeech \cite{Yang_SimpleSpeech_InterSpeech2024}} & 
\multirow{3}{*}{LibriTTS test-clean following \cite{Gao_E3TTS_ASRU2024} (1 utt/spk)} 
&
\underline{\resizebox{1.5cm}{!}{Nat (NMOS) $\uparrow$}}: SimpleSpeech (4.45) $>$ XTTS (4.19)  $>$ VALL-E X (4.07) \\ 
& & 
\underline{\resizebox{1.5cm}{!}{Qual (MCD) $\downarrow$}}: SimpleSpeech (7.51) $>$ VALL-E X (8.16) $>$ XTTS (9.93)\\
& &
\underline{\resizebox{1.5cm}{!}{Qual (WER) $\downarrow$}}: SimpleSpeech (3.30) $>$ XTTS (5.60) $=$ VALL-E X (5.60) \\
\midrule

\multirow{2}{*}{DINO-VITS \cite{Pankov_DINO_VITS_InterSpeech2024}} 
&
\multirow{2}{*}{CHiME3 (15 reference-audio/spk, 8 spks)} 
&
\underline{\resizebox{1.5cm}{!}{Nat (NMOS) $\uparrow$}}: [Clean] DINO-VITS (4.00) $>$ YTTS (3.96)\\ 
& & 
\underline{\resizebox{1.5cm}{!}{Nat (NMOS) $\uparrow$}}: [Noisy] DINO-VITS (3.55) $>$ YTTS+denoiser (3.28) $>$ YTTS (3.11) \\

\midrule

\multirow{2}{*}{Zhou et al. \cite{Zhou_InterSpeech2024}} 
&
\multirow{2}{*}{LibriTTS (test-clean \& test-other)| *WavLM and Hubert uses K-means (K=1024)} 
&
\underline{\resizebox{1.5cm}{!}{Nat (NMOS) $\uparrow$}}: Proposed (w/ WavLM) (3.54) $>$ Proposed (w/ Hubert) (3.24) $>$ VALL-E (2.50) \\ 
& & 
\underline{\resizebox{1.5cm}{!}{Qual (WER) $\uparrow$}}: Proposed (w/ WavLM) (7.02) $>$ Proposed (w/ Hubert) (9.90) $>$ VALL-E (23.31) \\ 

\midrule

\multirow{2}{*}{TacoLM \cite{Song_TacoLM_InterSpeech2024}} 
&
\multirow{2}{*}{LibriSpeech test-clean (4-10 sec/sample/spk, 40 spks) = 2.2h [similar to VALL-E]} 
&
\underline{\resizebox{1.5cm}{!}{Nat (CMOS) $\uparrow$}}: TacoLM (0.12) $>$ VALL-E (reference)\\ 
& &
\underline{\resizebox{1.5cm}{!}{Qual (WER) $\downarrow$}}: TacoLM (5.96) $>$ VALL-E (6.25)\\

\midrule

\multirow{1}{*}{Small-E \cite{Lemerle_SmallE_InterSpeech2024}} 
& 
\multirow{1}{*}{LibriTTS test-clean (15 samples)} 
&
\underline{\resizebox{1.5cm}{!}{Nat (NMOS) $\uparrow$}}: MetaVoice (3.80) $>$ Small-E (3.16) $>$ YTTS (2.56)\\ 

\midrule

\multirow{2}{*}{MaskGCT \cite{Wang_MaskGCT_ICLR2025}} 
& 
\multirow{2}{*}{LibriSpeech test-clean} 
&
\underline{\resizebox{1.5cm}{!}{Nat (CMOS) $\uparrow$}}: NS3 (0.16) $>$ MaskGCT (0.10) $>$ GT (Ref) $>$ VoiceCraft (-0.33) $>$ VB (-0.41) $>$ VALL-E (-0.52) $>$ XTTS v2 (-0.98) \\ 
& & 
\underline{\resizebox{1.5cm}{!}{Qual (WER) $\downarrow$}}: NS3 (1.94) $>$ VB (2.03) $>$ MaskGCT (2.63) $>$ XTTS v2 (4.20) $>$ VoiceCraft (4.68) $>$ VALL-E (5.90)\\

\bottomrule

\end{tabular}}
\label{ZeroShot_Comparison_nat}
\end{table*}

\begin{table*}
\centering
\caption{Comparison of Zero-shot TTS algorithms covering speaker similarity (Sim). For multiple datasets, results are averaged unless stated otherwise. Papers are arranged in ascending order by publication date.}
\resizebox{18cm}{!}{
\begin{tabular}{l|l|l}

\toprule

\multicolumn{1}{c|}{\textbf{Paper}} & \multicolumn{1}{c|}{\textbf{Test set, additional details}} & \multicolumn{1}{c}{\textbf{TTS Evaluation}} \\ 

\midrule

\multirow{2}{*}{DiffVoice \cite{Liu_DiffVoice_ICASSP2023}} 
& 
\multirow{2}{*}{LibriTTS (5 utt/reference, 21 reference audios) | [*P=Prompt]} 
&
\underline{\resizebox{1.5cm}{!}{Sim (SMOS) $\uparrow$}}: DiffVoice-P (4.61) $>$ VITS (x-vector) (3.80) $>$ YTTS (3.51) $>$ FS2 (x-vector) (3.44) $>$ MSS (3.15) \\
& &
\underline{\resizebox{1.5cm}{!}{Sim (SECS) $\uparrow$}}: DiffVoice-P (0.85) $>$ YTTS (0.81) $>$ VITS (x-vector) (0.80) $>$ MSS (0.76) $>$  FS2 (x-vector) (0.75)\\

\midrule

\multirow{2}{*}{GZS-TV \cite{Wang_GZS_TV_InterSpeech2023}} 
&
\multirow{2}{*}{LibriTTS test-set (5 utt/spk, 37 spks); VCTK (2 utt/spk, 108 spks)} 
&
\underline{\resizebox{1.5cm}{!}{Sim (SMOS) $\uparrow$}}: GZS-TV (3.79) $>$ YTTS (3.59) $>$ MSS (3.31) $>$ SS (3.30) \\
& &
\underline{\resizebox{1.5cm}{!}{Sim (SECS) $\uparrow$}}: GZS-TV (0.81) $>$ YTTS (0.79) $>$ MSS (0.75) $=$ SS (0.75) \\

\midrule

\multirow{1}{*}{STEN-TTS \cite{Tran_STEN_TTS_InterSpeech2023}} 
&
\multirow{1}{*}{VCTK (110 spks)} 
&
\underline{\resizebox{1.5cm}{!}{Sim (SMOS) $\uparrow$}}: STEN-TTS (3.72) $>$ SS (2.55) $>$ DiffSinger \cite{Liu_DiffSinger_AAAI2022} (2.50)\\

\midrule

\multirow{2}{*}{HierSpeech++ (HS++) \cite{Lee_HierSpeechPlusPlus_arxiv2023}} 
&
\multirow{1}{*}{LibriTTS test-clean (4837 samples)} 
& 
\underline{\resizebox{1.5cm}{!}{Sim (SMOS) $\uparrow$}}: HS++ (3.68) $>$ XTTS (3.49)  $>$ HS (3.31) $>$ VALL-E X (3.21) $>$ YTTS (3.06)\\
& 
\multirow{1}{*}{LibriTTS test-other (5120 samples)} 
& 
\underline{\resizebox{1.5cm}{!}{Sim (SECS) $\uparrow$}}: HS++ (0.89) $>$ VALL-E X (0.85) $>$ HS (0.80) $=$ YTTS (0.80) $>$ XTTS (0.77) \\

\midrule

\multirow{2}{*}{Latent Filling \cite{Bae_LatentFilling_ICASSP2024}} 
&
\multirow{2}{*}{VCTK (11 spks)}  & 
\underline{\resizebox{1.5cm}{!}{Sim (SMOS) $\uparrow$}}: Proposed (3.34) $>$ YTTS (2.79) $>$ SC-GlowTTS (2.08) \\ 
& &
\underline{\resizebox{1.5cm}{!}{Qual (SECS) $\downarrow$}}: Proposed (0.84) $>$ YTTS (0.81) $>$ SC-GlowTTS (0.60) \\

\midrule

\multirow{2}{*}{Mega-TTS 2 (MTTS2) \cite{jiang_MegaTTS2_ICLR2024}} 
& 
\multirow{2}{*}{LibriSpeech test-clean (100 sec/spk, 20 spks)} 
&
\underline{\resizebox{1.5cm}{!}{Sim (SMOS) $\uparrow$}}: MTTS2-300s (4.01) $>$ MTTS2-3s (3.75) $>$ VALL-E-3s (3.70) $>$ VALL-E-20s (3.25) \\
& & 
\underline{\resizebox{1.5cm}{!}{Sim (SECS) $\uparrow$}}: MTTS2-300s (0.93) $>$ MTTS2-3s (0.90) $>$ VALL-E-3s (0.89) $>$ VALL-E-20s (0.81) \\

\midrule

\multirow{2}{*}{USAT \cite{Wang_USAT_TASLP2024}} 
&
\multirow{2}{*}{ESLTTS, VCTK, LibriTTS test set (10 utt/spk, 30 spks) each} 
&
\multirow{1}{*}{\underline{\resizebox{1.5cm}{!}{Sim (SECS) $\uparrow$}}: USAT (0.73) $>$ YTTS (0.71) $>$ SS (0.65) $=$ MSS (0.65)} \\
& & 
\multirow{1}{*}{\underline{\resizebox{1.5cm}{!}{Sim (SMOS) $\uparrow$}}: USAT (3.69) $>$ YTTS (3.42) $>$ MSS (3.10) $>$ SS (3.06)}\\

\midrule

\multirow{2}{*}{MRMI-TTS \cite{Chen_MRMITTS_ACM_ALRLIP2024}} 
&
\multirow{2}{*}{VCTK (9 spks); LibriTTS test-clean} 
&
\underline{\resizebox{1.5cm}{!}{Sim (SMOS) $\uparrow$}}: MRMI-TTS (4.02) $>$ MSS (3.82) $>$ SS (3.78) $=$ FS2 (d-vector) (3.78) \\ 
& &
\underline{\resizebox{1.5cm}{!}{Sim (SECS) $\uparrow$}}: MRMI-TTS (0.82) $>$ MSS (0.81) $>$ SS (0.80) $=$ FS2 (d-vector) (0.80) \\

\midrule

\multirow{1}{*}{P-Flow \cite{Kim_PFlow_NIPS2024}} 
&
\multirow{1}{*}{LibriSpeech test-clean (4-10 sec/sample/spk, 40 spks) = 2.2h [similar to VALL-E]} 
&
\multirow{1}{*}{\underline{\resizebox{1.5cm}{!}{Sim (SECS) $\uparrow$}}: VALL-E (0.58) $>$ P-Flow (0.54) $>$ YTTS (0.34)}\\

\midrule

\multirow{1}{*}{StyleTTS 2 (STTS2) \cite{Li_StyleTTS2_NIPS2024}} 
& 
\multirow{1}{*}{LibriTTS test-clean (500 samples)} 
&
\underline{\resizebox{1.5cm}{!}{Sim (SMOS) $\uparrow$}}: STTS2 (4.03) $>$ STTS+HiFi-GAN (4.01) $>$ VITS (3.54) $>$ YTTS (2.42) \\

\midrule

\multirow{2}{*}{VALL-E 2 \cite{VALLE2_arxiv2024}} 
&
\multirow{2}{*}{LibriSpeech test-clean (4-10 sec/sample/spk, 40 spks) = 2.2h | *GroupSize = 1} 
& 
\underline{\resizebox{1.5cm}{!}{Sim (SECS) $\uparrow$}}: VALL-E 2 [FTS] (0.69) $>$ VALL-E [FTS] (0.68) $>$ VALL-E 2 (0.64) $>$ VALL-E (0.63) \\
& &
\underline{\resizebox{1.5cm}{!}{Sim (SMOS) $\uparrow$}}: VALL-E 2 (4.61) $>$ VALL-E (4.45) \\ 
\midrule

\multirow{1}{*}{CM-TTS \cite{CM_TTS_NAACL2024}} 
&
\multirow{1}{*}{VCTK (512 samples); LJSpeech (512 samples)} 
&
\multirow{1}{*}{\underline{\resizebox{1.5cm}{!}{Sim (SECS) $\uparrow$}}: CM-TTS (T=4) (0.72) $>$ DiffGAN-TTS \cite{Liu_DiffGAN_arxiv2022} (T=4) (0.71)} \\ 

\midrule

\multirow{2}{*}{NaturalSpeech 3 (NS3) \cite{Ju_NaturalSpeech3_ICML2024}} 
& 
\multirow{1}{*}{LibriSpeech test-clean (1 utt/spk, 40 spks)} 
&
\underline{\resizebox{1.5cm}{!}{Sim (SMOS) $\uparrow$}}: NS3 (4.01) $>$ NS2 (3.65) $>$ MTTS2 (3.63) $>$ VB (3.52) $>$ VALL-E (3.46) \\
& 
\multirow{1}{*}{*All are trained using LibriLight dataset} 
& 
\underline{\resizebox{1.5cm}{!}{Sim (SECS) $\uparrow$}}: NS3 (0.67) $>$ NS2 (0.55) $>$ MTTS2 (0.53) $>$ VB (0.48) $>$ VALL-E (0.47) \\

\midrule

\multirow{1}{*}{SpeechX \cite{SpeechX_TASLP2024}} 
&
\multirow{1}{*}{LibriSpeech test-clean (4-10 sec/sample/spk, 40 spks) = 2.2h [similar to VALL-E]} 
&
\multirow{1}{*}{\underline{\resizebox{1.5cm}{!}{Sim (SECS) $\uparrow$}}: SpeechX (VALL-E init) (0.58) $>$ SpeechX (Random init) (0.57) $=$ VALL-E (0.57)}\\

\midrule

\multirow{2}{*}{VoiceCraft \cite{VoiceCraft_ACL2024}} 
& 
\multirow{1}{*}{LibriTTS (125 utt); YouTube-GigaSpeech test-set (125 utt)} 
&
\multirow{1}{*}{\underline{\resizebox{1.5cm}{!}{Sim (SMOS) $\uparrow$}}: VoiceCraft (4.34) $>$ VALL-E (4.07) $>$ XTTS v2 \cite{Eren_Coqui2021} (3.44) $>$ YTTS (2.79)} \\
& 
\multirow{1}{*}{Objective|Subjective: 125|40 utt/dataset}  
&
\multirow{1}{*}{\underline{\resizebox{1.5cm}{!}{Sim (SECS) $\uparrow$}}: VoiceCraft (0.55) $>$ VALL-E (0.50) $>$ XTTS v2 \cite{Eren_Coqui2021} (0.47) $>$ YTTS (0.41)} \\

\midrule

\multirow{2}{*}{Xue et al. \cite{Xue_InterSpeech2024}} 
& 
\multirow{1}{*}{LibriSpeech test-clean (1249 sentences)} 
&
\underline{\resizebox{1.5cm}{!}{Sim (SMOS) $\uparrow$}}: Proposed (3.86) $>$ XTTS (3.44) $>$ VALL-E (3.29)\\
& 
\multirow{1}{*}{IEMOCAP (1079 sentences)} 
&
\underline{\resizebox{1.5cm}{!}{Sim (SECS) $\uparrow$}}: Proposed (0.76) $>$ XTTS (0.71) $>$ VALL-E (0.68)\\

\midrule

\multirow{2}{*}{Lee et al. \cite{Lee_InterSpeech2024}} 
&
\multirow{2}{*}{LibriTTS test-clean (500 sentences)} 
&
\underline{\resizebox{1.5cm}{!}{Sim (SMOS) $\uparrow$}}: Proposed (3.64) $>$ VITS (ECAPA TDNN) (3.23) $>$ VALL-E X (3.08)\\
& &
\underline{\resizebox{1.5cm}{!}{Sim (SECS) $\uparrow$}}: Proposed (0.51) $>$ VALL-E X (0.43) $>$ VITS (ECAPA TDNN) (0.39) \\

\midrule

\multirow{2}{*}{SimpleSpeech \cite{Yang_SimpleSpeech_InterSpeech2024}} & 
\multirow{2}{*}{LibriTTS test-clean following \cite{Gao_E3TTS_ASRU2024} (1 utt/spk)} 
&
\underline{\resizebox{1.5cm}{!}{Sim (SMOS) $\uparrow$}}: VALL-E X (4.15) $>$ SimpleSpeech (4.14) $>$ XTTS (4.00) \\
& & 
\underline{\resizebox{1.5cm}{!}{Sim (SECS) $\uparrow$}}: SimpleSpeech (0.96) $>$ XTTS (0.95) $=$ VALL-E X (0.95) \\

\midrule

\multirow{2}{*}{DINO-VITS \cite{Pankov_DINO_VITS_InterSpeech2024}} 
&
\multirow{2}{*}{CHiME3 (15 reference-audio/spk, 8 spks)} 
&
\underline{\resizebox{1.5cm}{!}{Sim (SMOS) $\uparrow$}}: [Clean] DINO-VITS (3.85) $>$ YTTS (3.33)\\
& &
\underline{\resizebox{1.5cm}{!}{Sim (SMOS) $\uparrow$}}: [Noisy] DINO-VITS (3.52) $>$ YTTS+denoiser (3.35) $>$ YTTS (3.20) \\

\midrule

\multirow{2}{*}{TacoLM \cite{Song_TacoLM_InterSpeech2024}} 
&
\multirow{2}{*}{LibriSpeech test-clean (4-10 sec/sample/spk, 40 spks) = 2.2h [similar to VALL-E]} 
&
\underline{\resizebox{1.5cm}{!}{Sim (SMOS) $\uparrow$}}: TacoLM (3.75) $>$ VALL-E (3.50)\\ 
& &
\underline{\resizebox{1.5cm}{!}{Sim (SECS) $\uparrow$}}: TacoLM (0.87) $>$ VALL-E (0.86)\\

\midrule

\multirow{1}{*}{Zhou et al. \cite{Zhou_InterSpeech2024}} 
&
\multirow{1}{*}{LibriTTS (test-clean \& test-other) | *WavLM and Hubert uses K-means (K=1024)} 
&
\multirow{1}{*}{\underline{\resizebox{1.5cm}{!}{Sim (SMOS) $\uparrow$}}: Proposed (w/ WavLM) (3.75) $>$ Proposed (w/ Hubert) (3.46) $>$ VALL-E (3.11)} \\ 

\midrule

\multirow{1}{*}{Small-E \cite{Lemerle_SmallE_InterSpeech2024}} 
& 
\multirow{1}{*}{LibriTTS test-clean (15 samples)} 
&
\underline{\resizebox{1.5cm}{!}{Sim (SMOS) $\uparrow$}}: MetaVoice (3.91) $>$ Small-E (3.08) $>$ YTTS (2.54)\\

\midrule

\multirow{2}{*}{MaskGCT \cite{Wang_MaskGCT_ICLR2025}} 
& 
\multirow{2}{*}{LibriSpeech test-clean} 
&
\underline{\resizebox{1.5cm}{!}{Sim (SMOS) $\uparrow$}}: MaskGCT (4.27) $>$ NS3 (4.26) $>$ VB (3.80) $>$ VoiceCraft (3.52) $>$ VALL-E (3.47) $>$ XTTS v2 (3.02) \\ 
& & 
\underline{\resizebox{1.5cm}{!}{Sim (SECS) $\uparrow$}}: MaskGCT (0.69) $>$ NS3 (0.67) $>$ VB (0.64) $>$ XTTS v2 (0.51) $>$ VALL-E (0.50) $>$ VoiceCraft (0.50) \\

\bottomrule

\end{tabular}}
\label{ZeroShot_Comparison_Similarity}
\end{table*}

On a more general note, many researchers target improving speaker representation to enhance voice cloning performance. Lee et al. \cite{Lee_ATVAE_ICNIDC2021} proposed a Tacotron2-based ZS-TTS system utilizing two types of speaker embeddings. The first is extracted from a speaker encoder, while the second is obtained using a proposed attention-based VAE from an embedding dictionary. Multi-spectroGAN \cite{Lee_MultiSpectrogan_AAAI2021} is a multi-speaker TTS model fully trained through adversarial training. It employs an adversarial style combination to improve the generalization of unseen speaker styles. The model consists of a FastSpeech2-based generator and a frame-level conditional discriminator, which enhances speaker embeddings by distinguishing real mel spectrograms from generated ones. Choi et al. \cite{Choi_APSIPA2022} proposed an adversarial speaker-consistency learning methodology to tackle the domain shift when generating speech for unseen speakers. This method utilizes external untranscribed speech during adversarial training to expose the model to a larger speaker pool. It is built on top of VITS architecture and demonstrated a performance increase in speech naturalness and speaker similarity. Schnell et al. \cite{Schnell_APW_SC2022} proposed an all-pass wrap (APW) implementation, commonly used for vocal tract length normalization, capable of performing zero-shot speech synthesis. The idea is that APW is capable of generalizing to various speakers in its original task; thus, introducing an APW layer to an encoder-decoder TTS architecture should enhance the zero-shot adaptation, as was demonstrated in its result with an increase in speaker similarity while decreasing the audio quality. Lyu et al. \cite{Lyu_SALN_ICIST2022} proposed a multi-scale zero-shot style transfer framework for ZS-TTS. They introduced a reference encoder with speaker-adaptive linear modulation (SALM), inspired by FiLM \cite{Perez_Film_AAAI2018}, to generate a scale and bias vector, improving adaptation performance to unseen speakers. They also designed a prosody extractor and predictor that leverages the reference audio and text to increase the diversity of the synthesized speech. Gorodetskii et al. \cite{Gorodetskii_arxiv2022} proposed a novel attention-based TTS capable of replicating target voice and generalizable to long utterances. This architecture includes a speaker encoder, a synthesizer, and a universal vocoder. They proposed the usage of dynamic convolution attention with a modified tacotron2-based synthesizer. In addition, they conditioned both the synthesizer and vocoder on a pre-trained speaker encoder. Lee et al. \cite{Lee_PVAETTS_ICASSP2022} introduced a progressive VAE (PVAE) that learns speaking style gradually and extends it to PVAE-TTS, a novel multi-speaker adaptive TTS that progressively learns normalized representation to adapt to new speakers with limited data. It also uses dynamic style layer normalization (DSLN) to improve style adaptation quality. Cory et al. \cite{Cory_COMPSAC2022} presented a novel speaker encoder that relies on an encoder coupled with a multi-scale approach to learn the local and global features of the target speaker. They used SC-GlowTTS transformer variation as their baseline, and they demonstrated that their method outperforms a modified GE2E speaker encoder with angular prototypical loss. WavThruVec \cite{Siuzdak_WavThruVec_InterSpeech2022} is a two-stage architecture consisting of an encoder (text2vec) that converts text into Wav2Vec 2.0 embeddings using a pre-trained Wav2Vec 2.0 model and a decoder (vec2wav) to obtain a generated waveform. These contributions improved the generalizability of unseen speakers. Kim \& Jeong et al. \cite{Kim_TransferLearning_InterSpeech2022} proposed a TTS transfer learning methodology, using VITS as a baseline, that employs Wav2Vec 2.0 representations to utilize a large amount of unlabeled data for TTS model pre-training. AdaSpeech 4 \cite{Wu_AdaSpeech4_InterSpeech2022} builds on top of AdaSpeech by introducing three major contributions: factorizing speaker characteristics as basis vectors and combining them through attention, employing CLN on the extracted speaker representation, and proposing a loss based on basis vector distribution. These contributions led to an improvement in speech quality and speaker similarity in a ZS-TTS scenario. Eskimez et al. \cite{Eskimez_InterSpeech2024} introduced a novel duration model, called total-duration aware (TDA), utilizing both text input and the total target duration. The purpose of this module is to maintain speech quality across various speech rate configurations while providing a more precise speech duration, leading to a performance boost in intelligibility and preserving the speaker's voice characteristics. VALL-T \cite{VALL_T_arxiv2024} is a generative transducer model designed to address robustness issues, such as hallucinations and high word error rates, observed in decoder-only TTS algorithms like VALL-E. These algorithms suffer from the absence of explicit duration modeling, which results in poor performance due to the lack of monotonic alignment constraints. VALL-T introduced shifting relative position embeddings for the input phonemes, enforcing monotonic alignment constraints and enhancing TTS robustness. On the other hand, other works focus on speaker conditioning techniques to enable and improve the performance of ZS-TTS. Choi et al. \cite{Choi_SNAC_SPL2022} introduced a speaker-normalized affine coupling (SNAC) layer to enable ZS-TTS through conditioning techniques. It normalizes the input through the predicted parameters from speaker embeddings. SC-CNN \cite{Yoon_SC_CNN_SPL2023} is a speaker-conditioning technique that predicts convolutional kernels from speaker embeddings and applies 1D convolution with these kernels to phoneme sequences. Chen et al. \cite{Chen_EURASIP2024} proposed a two-branch speaker control module (TSCM) designed to enhance speaker similarity in few-shot and zero-shot TTS. This module includes a gated convolutional network (GCN) for local feature modeling and a gated recurrent unit (GRU) to obtain utterance-level features. The authors demonstrate the module's effectiveness through integration with FastSpeech 2 and VITS.

More recently, ZS-TTS systems started integrating diffusion models for high-quality speech synthesis. Jiang et al. \cite{Jiang_InterSpeech2024} introduced a ZS-TTS model focusing on timbre generalization and prosody modeling. This includes a global vector to model speaker timbre while guiding prosody modeling. In addition, they adopted a DDPM-based pitch predictor and a prosody adapter for hierarchical prosody modeling. Furthermore, Guided-TTS 2 \cite{Kim_GuidedTTS2_arxiv2022}, previously discussed for its few-shot TTS generation, is also capable of zero-shot TTS through a pre-trained speaker encoder. Bang et al. \cite{Bang_Sensors2023} proposed a ZS-TTS system by building on top of the diffusion-based Grad-TTS \cite{Popov_GradTTS_ICML2021} and enabling speech synthesis for unseen speakers. They utilize a pre-trained speaker recognition model to obtain speaker information from reference speech and expand it through information perturbation. Similarly, Grad-StyleSpeech \cite{Kang_GradStyleSpeech_ICASSP2023} is a ZS-TTS model that utilizes a hierarchical transformer encoder and score-based diffusion model, proposed in Grad-TTS, for generating speech while replicating the target speaker voice. U-Style \cite{Li_UStyle_TASLP2024} is a Grad-TTS-based ZS-TTS algorithm, enabling control over speech synthesis by utilizing independent speaker and style audio references from unseen speakers. It effectively uses a cascaded U-Net structure and signal perturbation to disentangle speaker-style information. Additionally, it employs normalization techniques and a diffusion-based decoder to enhance the quality and naturalness of synthesized speech. DiffVoice \cite{Liu_DiffVoice_ICASSP2023} is a novel TTS algorithm that starts with encoding speech into phoneme-level representation using VAE and is enhanced using adversarial training. It then jointly models these representations and phoneme durations with a diffusion model. More recently, NaturalSpeech 2 \cite{Shen_NaturalSpeech2_ICLR2024} tackles the challenge of general large-scale TTS systems' robustness, including unstable prosody, skipping/repeating, and voice quality. It uses a neural audio codec with a residual vector quantizer and a diffusion model to generate latent vectors conditioned on text input, and the decoder uses these vectors to generate the target speech. RALL-E \cite{RALLE_arxiv2024} introduced chain-of-thought (CoT) prompting within an NCLM to improve robustness in LLM-based TTS. This means breaking a complex task into a series of more straightforward ones to improve performance. This model starts by predicting prosody tokens (pitch and duration). Then, it predicts speech tokens conditioned on those and the input sequence. In addition, it utilizes the predicted duration to mask irrelevant phonemes and prosody tokens to ensure the model concentrates only on the relevant parts. Zhou et al. \cite{Zhou_InterSpeech2024} proposed a phonetic-enhanced language modeling method by leveraging SSL to train the AR language model. Subsequently, the mapping from phonetic variations to fine-grained acoustic details occurs in an NAR manner. This approach reduces error propagation during training, improving robustness and naturalness. VALL-E 2 \cite{VALLE2_arxiv2024} introduced two VALL-E enhancements achieving human parity. The first is repetition-aware sampling, which enhances the stability of the decoding process compared to random sampling. Additionally, five-time sampling (FTS) generates five outputs and selects the best one based on specific metrics, improving robustness. The second modification, grouped code modeling, increases inference speed by organizing codec codes into groups.

As TTS algorithms develop, researchers focus on exploring algorithms and their components. Cooper et al. \cite{Cooper_ICASSP2020} explored the effect of various neural speaker embeddings on speaker similarity performance of unseen speakers. They showed that LDE with angular softmax loss performs better than x-vectors in speaker verification. SATTS \cite{Goswami_SATTS_InterSpeech2022} explored speaker attractors, a speech separation technique, for ZS-TTS algorithms. Speaker attractors are embedding vectors that pull time-frequency bins of the same speaker together while pushing those of other speakers, which proved useful with worse recording conditions. Recently, Azizah et al. \cite{Azizah_IEEEACCESS2024} explored 24 Tacotron2-based zero-shot TTS variants and observed their performance on speech synthesis for dysphonia speakers. These variants include changing the type of input sequence, speaker embedding and its position, and loss function. They concluded that the best-performing models use a grapheme-phoneme input sequence, a speaker model to generate speaker embeddings placed at the TTS encoder only, and a combined loss using speaker consistency and frame-level speech loss.

\textbf{Non-English ZS-TTS} Zero-shot TTS algorithms have shown remarkable advancements in synthesizing human-like speech. However, it is important to focus on other languages to address the world's linguistic diversity. Papers \cite{Zhang_iEmoTTS_TASLP2023, Zhou_CDFSE_InterSpeech2022 ,Li_SponTTS_ICASSP2024} focus on Mandarin, paper \cite{Fujita_ICASSP2024} focus on Japanese, while \cite{Dat_Calib_StyleSpeech_ICAIBDDE2021, Ngoc_AdaptTTS_CSS2023} focus on the Vietnamese language. Zhou et al. \cite{Zhou_CDFSE_InterSpeech2022} proposed a content-dependent fine-grained speaker embedding (CDFSE) approach tackling Mandarin ZS-TTS. They extract local content and speaker embeddings from a reference speech, introducing pronunciation information. This approach improves speaker similarity for unseen speakers by leveraging a content-dependent reference attention module. iEmoTTS \cite{Zhang_iEmoTTS_TASLP2023} is a novel TTS system for cross-speaker emotion transfer through prosody-timbre disentanglement. This paper proposed a system consisting of an emotion encoder, a prosody predictor, a timbre encoder, and a probability-based emotion intensity control mechanism. These contributions led to speech synthesis while controlling emotion intensity for new speakers. SponTTS \cite{Li_SponTTS_ICASSP2024} is a two-stage approach, based on neural bottleneck (BN) features, for spontaneous TTS generation used in casual conversations. The first stage, text2BN, based on FastSpeech, maps text to speaker-independent BN features conditioned on spontaneous style representation using a conditional VAE (CVAE). The second stage, BN2Wave, following a VITS-based model, transforms BN intermediate values conditioned on speaker embeddings to generate speech similar to the target speaker. For Japanese, Fujita et al. \cite{Fujita_ICASSP2024} introduced a zero-shot TTS method, incorporating adapters into an SSL model fine-tuned using noisy reference speech. They also included a speech enhancement module for further noise reduction. For Vietnamese, Calib-StyleSpeech \cite{Dat_Calib_StyleSpeech_ICAIBDDE2021} is a FastSpeech2-based ZS-TTS algorithm that tackles accent imbalance between datasets. Their model includes a style extraction block to condition the acoustic features. Simultaneously, an independent block extracts a content vector to disentangle content-style information through a mutual information constraint. Adapt-TTS \cite{Ngoc_AdaptTTS_CSS2023} is a zero-shot TTS model that uses an extracting mel-vector (EMV) to extract speaker and style information and a diffusion-based mel spectrogram denoiser to improve the quality of the generated voice.

\begin{figure*}
\centering
  \includegraphics[width=0.96\linewidth]{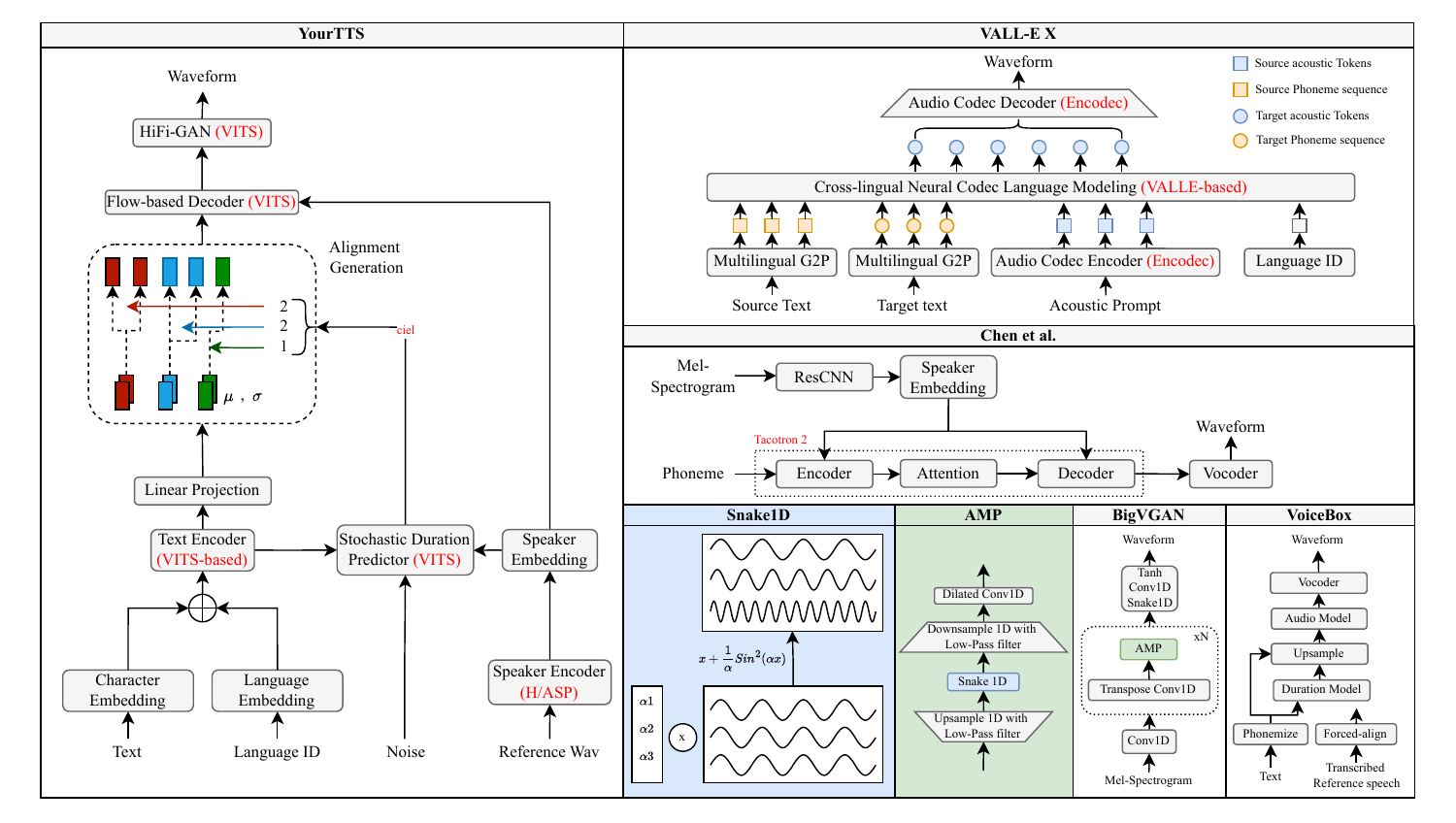}
\caption{Multilingual Voice Cloning algorithms, adapted from YourTTS \cite{Casanova_YourTTS_ICML2022}, VALL-E X \cite{Zhang_VALL_E_X_arxiv2023}, BigVGAN \cite{Lee_BigVGAN_ICLR2023}, VoiceBox \cite{Le_Voicebox_NIPS2024}, and Chen et al. \cite{Chen_InterSpeech2019}.}
\label{multilingual_TTS_Figure1}  
\end{figure*}

\textbf{Summary:} Zero-Shot TTS algorithms witnessed the most significant increase in papers within voice cloning in the past few years. After the release of SPEAR-TTS and VALL-E, many works started employing codec-based architecture for zero-shot TTS. Tables \ref{ZeroShot_Comparison_nat} and  \ref{ZeroShot_Comparison_Similarity} show the performance of many zero-shot TTS architectures in naturalness, speech quality, and speaker similarity. Similar to FS-TTS, calculating WER and CER require using a pre-trained ASR to evaluate the speech quality of the generated waveform. These systems include Whisper ASR (base or large), CTC-based HuBERT \cite{Hsu_Hubert_TASLP2021}, Conformer Transducer \cite{Gulati_Conformer_Interspeech2020}, and Wav2Vec 2.0 \cite{Baevski_Wav2Vec_NIPS2020} to quantify the speech quality performance of TTS systems. Speaker Embedding Cosine similarity scores are obtained through the Resemblyzer package, WavLM \cite{Chen_WavLM_JSTSP2022}, XLSR-53 \cite{Conneau_XLSR53_InterSpeech2021}, TitaNet-L, and SpeechBrain toolkit \cite{Ravanelli_ConverasationalSpeechBrain_arxiv2024, Ravanelli_SpeechBrain_arxiv2021}.

\subsection{Multilingual Voice Cloning}

Building multilingual TTS while maintaining speaker characteristics is crucial for advancing personalized chatbots. We will explore speaker adaptation, FS-TTS, and ZS-TTS across multiple languages. Fig. \ref{multilingual_TTS_Figure1} shows some of these algorithms, while Table \ref{Multi-lingual} lists work focusing on such systems.

\textbf{2 Languages:} This paragraph will cover multilingual TTS systems with only two languages in the following order: speaker adaptation, FS-TTS, and ZS-TTS systems. Xin et al. \cite{Xin_InterSpeech2021} proposed a speaker adaptation method for mono/cross-lingual speaker adaptation for English and Japanese. They train a language-independent speaker encoder and use the resultant speaker embeddings to train a monolingual multi-speaker TTS. Finally, they adapt to new speakers by fine-tuning the model with a speaker consistency loss. Other works focus on multilingual FS-TTS algorithms, Chen et al. \cite{Chen_InterSpeech2019} introduced a novel speaker embedding network that enables bilingual learning between English and Chinese, even for unseen speakers. This consists of a ResCNN \cite{Li_DeepSpeaker_arxiv2017} speaker encoder, a Tacotron2-based mel spectrogram generation network, and a Griffin-Lim vocoder. AdaDurIAN \cite{Zhang_adadurian_arxiv2020} improved the DurIAN-based average model and adapted it into a few-shot mono/cross-lingual speaker adaptation through a few minutes of monolingual data. The authors' contributions include building a speaker-independent content encoder, fine-tuning speaker embedding and decoder instead of the entire model for better pronunciation, and adopting an LSTM post-net module to improve speech quality. FastGraphTTS \cite{Wang_FastGraphTTS_ICTAI2023} is an end-to-end TTS framework that incorporates syntactic information to improve prosody consistency between the input text and the generated speech using a graph neural network (GNN). Other works focus on ZS-TTS algorithms, Zheng et al. \cite{Zheng_ICASSP2022} presented a multi-stream text encoder and efficient speaker representation to improve the performance for zero-shot cross-lingual TTS on unseen speakers. They utilize a speaker encoder with random sampling and a language adversarial loss for speaker-content-language disentanglement. NNSpeech \cite{Zhao_NNSpeech_ICASSP2022} is a speaker-guided CVAE that enables unseen speaker voice cloning with only one utterance. Cai et al. \cite{Cai_CSL2023} proposed a bilingual Mandarin-English TTS system operating in two distinct ways depending on the data available per speaker. In the case of limited data, they used cascading multi-speech modules, including a VC module for monolingual and cross-lingual speaker adaptation. VALL-E X \cite{Zhang_VALL_E_X_arxiv2023} is an extension of the VALL-E network to enable cross-lingual speech synthesis. It introduced training a multilingual conditional codec language modeling and synthesizing the unseen speaker's characteristics into the target language. More recently, Seed-TTS \cite{SeedTTS_arxiv2024} is a family of large-scale TTS models capable of performing ZS-TTS and emotion control through instruction fine-tuning. They use a self-distillation approach for timbre disentanglement, enabling high-quality VC and a reinforcement learning-based approach to enhance the model's robustness. Additionally, they introduced a diffusion-based NAR variant that performs speech generation in an end-to-end manner without pre-estimated phoneme durations. MobileSpeech \cite{Jiang_MobileSpeech_ACL2024} is a light-weight FastSpeech2-based ZS-TTS model that leverages a parallel speech mask codec decoder that uses hierarchical information from the speech codec and employs weight mechanisms across its layers during generation, achieving competitive performance in both English and Mandarin while reducing latency. Wang et al. \cite{Wang_MaskGCT_ICLR2025} proposed Masked Generative Codec Transformer (MaskGCT), a two-stage fully NAR TTS model that does not require an explicit phoneme duration predictor and text-speech alignment supervision. The first stage predicts semantic tokens using a speech SSL model, while the second stage, based on SoundStorm \cite{Borsos_SoundStorm_arxiv2023}, uses a masked generative codec transformer conditioned on the semantic tokens. Unlike previous methods, Latent Filling \cite{Bae_LatentFilling_ICASSP2024} is a speaker embedding latent space augmentation through a consistency loss to improve the performance of intra-lingual (English to English) and cross-lingual (Korean to English) ZS-TTS. This ensures that the generated acoustic features of the augmented speaker embedding maintain the same speaker representation.

\begin{table}[t]
\caption{Table of Language Acronyms.}
\centering
\resizebox{8.5cm}{!}{
\begin{tabular}{cc||cc||cc}
\toprule
\textbf{Acronym} & \textbf{Full Form} & \textbf{Acronym} & \textbf{Full Form} & \textbf{Acronym} & \textbf{Full Form} \\

\midrule

\multicolumn{1}{l|}{EN} & \multicolumn{1}{l||}{English} & 
\multicolumn{1}{l|}{PT} & \multicolumn{1}{l||}{Portuguese} & \multicolumn{1}{l|}{JV} & \multicolumn{1}{l}{Javanese} \\

\midrule

\multicolumn{1}{l|}{ZH} & \multicolumn{1}{l||}{Chinese} & 
\multicolumn{1}{l|}{SP} & \multicolumn{1}{l||}{Spanish} & 
\multicolumn{1}{l|}{VI} & \multicolumn{1}{l}{Vietnamese} \\

\midrule

\multicolumn{1}{l|}{KO} & \multicolumn{1}{l||}{Korean} & 
\multicolumn{1}{l|}{JA} & \multicolumn{1}{l||}{Japanese} & 
\multicolumn{1}{l|}{DE} & \multicolumn{1}{l}{German}\\

\midrule

\multicolumn{1}{l|}{SU} & \multicolumn{1}{l||}{Sundanese} &
\multicolumn{1}{l|}{ID} & \multicolumn{1}{l||}{Indonesian} & 
\multicolumn{1}{l|}{FR} & \multicolumn{1}{l}{French} \\

\midrule

\multicolumn{1}{l|}{HI} & \multicolumn{1}{l||}{Hindi} &
\multicolumn{1}{l|}{MR} & \multicolumn{1}{l||}{Marathi} & 
\multicolumn{1}{l|}{-} & \multicolumn{1}{l}{-}\\

\bottomrule
\end{tabular}}
\label{Language_Acronym}  
\end{table}

\begin{table}[t]
\centering
\caption{Multilingual Voice Cloning papers, following abbreviations from table \ref{Language_Acronym}.}
\resizebox{8.5cm}{!}{
\begin{tabular}{c|c||c|c}
\toprule
\textbf{Ref} & \textbf{Languages} & \textbf{Ref} & \textbf{Languages} \\ \midrule

\multicolumn{1}{l|}{\cite{Xin_InterSpeech2021}} &
\multicolumn{1}{l||}{EN-JA} &
\multicolumn{1}{l|}{\cite{Zhao_NNSpeech_ICASSP2022, Zhang_adadurian_arxiv2020, Wang_FastGraphTTS_ICTAI2023, Cai_CSL2023, Zheng_ICASSP2022, Zhang_VALL_E_X_arxiv2023, Chen_InterSpeech2019, SeedTTS_arxiv2024, Wang_MaskGCT_ICLR2025, Jiang_MobileSpeech_ACL2024}} &
\multicolumn{1}{l}{EN-ZH} \\

\midrule

\multicolumn{1}{l|}{\cite{Bae_LatentFilling_ICASSP2024}}  &
\multicolumn{1}{l||}{EN-KO} &
\multicolumn{1}{l|}{\cite{Casanova_YourTTS_ICML2022}} &
\multicolumn{1}{l}{EN-FR-PT} \\

\midrule

\multicolumn{1}{l|}{\cite{Shah_ParrotTTS_EACL2024}} &
\multicolumn{1}{l||}{HI-DE-SP-MR} & 
\multicolumn{1}{l|}{\cite{Azizah_IEEE_Access2022}} &
\multicolumn{1}{l}{EN-ID-JV-SU}\\

\midrule

\multicolumn{1}{l|}{\cite{Tran_STEN_TTS_InterSpeech2023}} &
\multicolumn{1}{l||}{EN-ZH-ID-JA-VI} & 
\multicolumn{1}{l|}{\cite{Hemati_arxiv2020}} &
\multicolumn{1}{l}{EN-DE-SP-FR-KO}\\

\midrule

\multicolumn{1}{l|}{\cite{Le_Voicebox_NIPS2024}} &
\multicolumn{1}{l||}{6 languages} & 
\multicolumn{1}{l|}{\cite{Gong_ZMM_TTS_TASLP2024}} &
\multicolumn{1}{l}{6 languages}\\

\midrule

\multicolumn{1}{l|}{\cite{Jeong_TASLP2024}} &
\multicolumn{1}{l||}{7 languages} & 
\multicolumn{1}{l|}{\cite{Jeon_EACL2024}} &
\multicolumn{1}{l}{8 languages} \\

\midrule

\multicolumn{1}{l|}{\cite{Lee_BigVGAN_ICLR2023}} &
\multicolumn{1}{l||}{9 languages} & 
\multicolumn{1}{l|}{\cite{Kim_CLaM_TTS_ICLR2024}} &
\multicolumn{1}{l}{11 languages} \\

\midrule

\multicolumn{1}{l|}{\cite{Lux_AACL_IJCNLP2022}} &
\multicolumn{1}{l||}{12 languages} & 
\multicolumn{1}{l|}{\cite{Casanova_XTTS_InterSpeech2024}} &
\multicolumn{1}{l}{16 languages} \\

\midrule

\multicolumn{1}{l|}{\cite{Yang_InterSpeech2020}} &
\multicolumn{1}{l||}{50 language locales} & 
\multicolumn{1}{l|}{-} &
\multicolumn{1}{l}{-} \\

\bottomrule

\end{tabular}}
\label{Multi-lingual}
\end{table}

\textbf{3 to 5 Languages:} Unlike TTS models with only two languages, more languages require a general algorithm as languages differ in their alphabets and pronunciation. We will start with limited multilingual algorithms. YourTTS \cite{Casanova_YourTTS_ICML2022} is an extension of VITS for zero-shot TTS and multilingual training. This includes using text instead of phonemes, as not all languages have suitable open-source grapheme-to-phoneme converters. It also integrates a trainable language embedding with input characters and a H/ASP speaker encoder \cite{Heo_HASP_arxiv2020} to enable multilingual zero-shot TTS. Parrot-TTS\cite{Shah_ParrotTTS_EACL2024} is a modularized TTS model that leverages SSL and performs well on adapting to unseen languages while preserving the speaker's voice characteristics, and it performs well for unseen speaker ZS-TTS in four languages. This architecture trains a seq2seq module as its text encoder and a pre-trained HuBERT as its speech encoder. The embeddings obtained from both generate the output waveform using a HiFi-GANv2 vocoder. Azizah et al. \cite{Azizah_IEEE_Access2022} proposed a novel partial network-based transfer learning to handle low-resource language. They utilized text and language encoders to improve multilingual performance. In addition, they used an explicit style control for TTS conditioning and an utterance-level reconstruction loss to improve the performance of zero-shot speaker adaptation. STEN-TTS \cite{Tran_STEN_TTS_InterSpeech2023} is a diffusion-based multilingual ZS-TTS capable of cross-lingual synthesis with a 3-second reference audio. This method also introduces style-enhanced normalization (STEN) into the diffusion framework to maintain the speaker's style, a challenge to previous diffusion-based TTS systems. Hemati et al. \cite{Hemati_arxiv2020} introduced a Tacotron2-based model suitable for multilingual speaker adaptation. Unlike previous methods, they use the International Phonetic Alphabet (IPA) as a direct input that goes through a trainable LUT for better generalization across languages. They explore transfer learning to enhance inter-lingual and cross-lingual speaker adaptation performance.

\textbf{5+ Languages:} This section discusses the broader realm of multilingual algorithms. Yang et al. \cite{Yang_InterSpeech2020} explored building a universal seq2seq TTS model capable of synthesizing any speaker's voice in any language. This framework includes a transformer-based acoustic model, a WaveNet vocoder, and global conditioning from speaker and language networks. The global conditioning uses speaker embeddings obtained from a LUT followed by a mapping network to generate features with similar dynamic range. Similarly, the language embeddings are obtained from a language network with the same architecture. To enable a large multilingual universal synthesis, the authors proposed concatenating these features with the output of the acoustic model. Lux et al. \cite{Lux_AACL_IJCNLP2022} proposed a modified encoder and the usage of language-agnostic meta-learning (LAML) for zero-shot cross-lingual TTS that performs well for unseen speakers. This work uses articulatory feature vectors \cite{Lux_ACL_2022} to share knowledge across languages, making it a language-agnostic approach. It also introduced the usage of word boundaries to improve multilingual TTS performance. Jeon et al. \cite{Jeon_EACL2024} explored language-agnostic voice cloning through a multi-level attention aggregation approach, evaluating its effectiveness on unseen speakers within seen languages and unseen speakers and languages. Voicebox \cite{Le_Voicebox_NIPS2024} is a versatile text-guided speech generative model capable of noise removal, content editing, diverse sample generation, style conversion, and mono/cross-lingual zero-shot TTS. It is a flow-matching model trained through infilling given a text and audio context. ZMM-TTS \cite{Gong_ZMM_TTS_TASLP2024} is a multilingual zero-shot TTS model that leverages discrete speech representation from an SSL model and is capable of ZS-TTS for unseen speakers and languages, including low-resource languages. The authors explored the effect of various input representations on multilingual TTS performance, such as characters, IPA, or pre-trained phoneme representation, in which the latter demonstrated better overall performance. CLAM-TTS \cite{Kim_CLaM_TTS_ICLR2024} is a codec-based TTS model that employs a probabilistic residual vector quantization to improve token length compression while performing with comparable or better performance compared to SOTA neural codec-based TTS models. Jeong et al. \cite{Jeong_TASLP2024} introduced a transfer learning scheme consisting of two stages: unsupervised pre-training using speech-only data by leveraging self-supervised speech representations obtained through Wav2Vec 2.0. Then, supervised fine-tuning to generalize to new languages and speakers efficiently. To handle multilingual diversity, they utilized XLSR-53 \cite{Babu_XLSR_InterSpeech2022}, based on Wav2Vec 2.0, to obtain language-independent linguistic features useful for various downstream tasks. XTTS \cite{Casanova_XTTS_InterSpeech2024} builds on top of Tortoise TTS \cite{Betker_tortoise_arxiv2023} to enable multilingual training. This model consists of VQ-VAE, an encoder comprising various components, GPT-2, and a HiFi-GAN vocoder. They use a BPE tokenizer \cite{BPE_arxiv2012} to obtain text tokens as input to GPT-2. They also romanized Korean, Japanese, and Chinese before tokenization. Researchers mainly focus on acoustic modeling, leading to less research on other components, such as the vocoder to handle speaker diversity. BigVGAN \cite{Lee_BigVGAN_ICLR2023} introduced a universal GAN-based vocoder that achieves SOTA performance on zero-shot out-of-distribution conditions, including unseen speakers and languages. They enhanced the GAN-based generator by introducing a periodic activation function that provides inductive bias into audio synthesis. It also introduced an anti-aliased multi-periodicity composition (AMP) module that composes multiple signals with learnable periodicities and a low-pass filter to reduce high-frequency artifacts.

\begin{table*}[t]
\centering
\caption{Speech datasets across various languages with a focus on TTS datasets.}
\resizebox{18cm}{!}{
\begin{tabular}{cccccc}
\toprule
\textbf{Dataset} & \textbf{Task} & \textbf{Amount [hours (h), utterances (utt)]} & \textbf{Speakers [Male, Female]} & \textbf{Sampling Rate} & \textbf{Languages}\\ \midrule

\rowcolor{lightgray}

\multicolumn{6}{c}{Single Language} \\

\midrule

\multicolumn{1}{l|}{CHiME3 \cite{Barker_CHiMEdataset_CSL2017}} &  
\multicolumn{1}{c|}{ASR} &  
\multicolumn{1}{c|}{11698 utt} &  
\multicolumn{1}{c|}{83 [6+, 6+]} &
\multicolumn{1}{c|}{16 kHz} &  
\multicolumn{1}{c}{English} \\
 
\midrule

\multicolumn{1}{l|}{LibriSpeech \cite{Panayotov_LibriSpeech_ICASSP2015}} &  
\multicolumn{1}{c|}{ASR} &  
\multicolumn{1}{c|}{982h} &  
\multicolumn{1}{c|}{2484 [1283, 1201]} &
\multicolumn{1}{c|}{16 kHz} &  
\multicolumn{1}{c}{English} \\

\midrule

\multicolumn{1}{l|}{GigaSpeech \cite{Chen_GigaSpeech_InterSpeech2021}} &  
\multicolumn{1}{c|}{ASR} &  
\multicolumn{1}{c|}{[10kh labeled, 30kh unlabeled/weakly labeled]} &  
\multicolumn{1}{c|}{N/A} &
\multicolumn{1}{c|}{16 kHz} &  
\multicolumn{1}{c}{English} \\
 
\midrule

\multicolumn{1}{l|}{VCTK \cite{Yamagishi_VCTK_Edinburgh2019}} & 
\multicolumn{1}{c|}{TTS} &  
\multicolumn{1}{c|}{44h} &  
\multicolumn{1}{c|}{109 [N/A, N/A]} &  
\multicolumn{1}{c|}{48 kHz} &  
\multicolumn{1}{c}{English} \\

\midrule

\multicolumn{1}{l|}{LJSpeech \cite{Ito_LJSpeech_online2017}} & 
\multicolumn{1}{c|}{TTS} &  
\multicolumn{1}{c|}{24h} &  
\multicolumn{1}{c|}{1 [0, 1]} &  
\multicolumn{1}{c|}{22.05 kHz} &  
\multicolumn{1}{c}{English} \\
 
\midrule

\multicolumn{1}{l|}{ESLTTS \cite{Wang_USAT_TASLP2024}} & 
\multicolumn{1}{c|}{TTS} &  
\multicolumn{1}{c|}{37h} &  
\multicolumn{1}{c|}{134 [N/A, N/A]} &  
\multicolumn{1}{c|}{24.00 kHz} &  
\multicolumn{1}{c}{English} \\
 
\midrule

\multicolumn{1}{l|}{Blizzard 2011 \cite{King_blizzard2011}} & 
\multicolumn{1}{c|}{TTS} &  
\multicolumn{1}{c|}{16.6h} &  
\multicolumn{1}{c|}{1 [0, 1]} &  
\multicolumn{1}{c|}{16 kHz} &  
\multicolumn{1}{c}{English} \\

\midrule

\multicolumn{1}{l|}{Blizzard 2013 \cite{King_blizzard2013}} & 
\multicolumn{1}{c|}{TTS} &  
\multicolumn{1}{c|}{319h} &  
\multicolumn{1}{c|}{1 [0, 1]} &  
\multicolumn{1}{c|}{Varied} &  
\multicolumn{1}{c}{English} \\
 
\midrule

\multicolumn{1}{l|}{Hi-Fi TTS \cite{Bakhturina_HiFiTTS_InterSpeech2021}} &  
\multicolumn{1}{c|}{TTS} &  
\multicolumn{1}{c|}{292h} &  
\multicolumn{1}{c|}{10 [4, 6]} &  
\multicolumn{1}{c|}{44.1 kHz} &  
\multicolumn{1}{c}{English} \\

\midrule

\multicolumn{1}{l|}{LibriLight \cite{Kahn_LibriLight_ICASSP2020}} & 
\multicolumn{1}{c|}{ASR} &  
\multicolumn{1}{c|}{[10h labeled, 60kh unlabeled]} &  
\multicolumn{1}{c|}{7000+ [N/A, N/A]} &  
\multicolumn{1}{c|}{16 kHz} &  
\multicolumn{1}{c}{English} \\

\midrule

\multicolumn{1}{l|}{LibriTTS \cite{Zen_LibriTTS_arxiv2019}} &  
\multicolumn{1}{c|}{TTS} &  
\multicolumn{1}{c|}{586h} &  
\multicolumn{1}{c|}{2456 [1271, 1185]} &  
\multicolumn{1}{c|}{24 kHz} &  
\multicolumn{1}{c}{English} \\

\midrule

\multicolumn{1}{l|}{JSUT \cite{Sonobe_JSUT_arxiv2017}} &  
\multicolumn{1}{c|}{TTS} &  
\multicolumn{1}{c|}{10h} &  
\multicolumn{1}{c|}{1 [0, 1]} &
\multicolumn{1}{c|}{48 kHz} &  
\multicolumn{1}{c}{Japanese} \\

\midrule

\multicolumn{1}{l|}{JVS \cite{Takamichi_JVS_arxiv2019}} &  
\multicolumn{1}{c|}{VC, TTS} &  
\multicolumn{1}{c|}{30h} &  
\multicolumn{1}{c|}{100 [49, 51]} &
\multicolumn{1}{c|}{24 kHz} &  
\multicolumn{1}{c}{Japanese} \\

\midrule

\multicolumn{1}{l|}{CSMSC \cite{CSMSC_DataBaker_2017}} &  
\multicolumn{1}{c|}{TTS} &  
\multicolumn{1}{c|}{12h} &  
\multicolumn{1}{c|}{1 [0, 1]} &
\multicolumn{1}{c|}{48 kHz} &  
\multicolumn{1}{c}{Mandarin} \\
 
\midrule

\multicolumn{1}{l|}{AISHELL-1 \cite{Bu_AISHELL1_O_COCOSDA2017}} & 
\multicolumn{1}{c|}{ASR} & 
\multicolumn{1}{c|}{170h} &  
\multicolumn{1}{c|}{400 [186, 214]} &  
\multicolumn{1}{c|}{16 kHz} &  
\multicolumn{1}{c}{Mandarin} \\
 
\midrule

\multicolumn{1}{l|}{AISHELL-2 \cite{Du_AISHELL2_arxiv2018}} & 
\multicolumn{1}{c|}{ASR} & 
\multicolumn{1}{c|}{1000h} &  
\multicolumn{1}{c|}{1991 [845, 1146]} &  
\multicolumn{1}{c|}{16 kHz, 44.1 kHz} &  
\multicolumn{1}{c}{Mandarin} \\
 
\midrule

\multicolumn{1}{l|}{AISHELL-3 \cite{Shi_AISHELL3_arxiv2020}} & 
\multicolumn{1}{c|}{TTS} &  
\multicolumn{1}{c|}{85h} &  
\multicolumn{1}{c|}{218 [42, 176]} &  
\multicolumn{1}{c|}{44.1 kHz} &  
\multicolumn{1}{c}{Mandarin} \\
 
\midrule

\multicolumn{1}{l|}{WENETSPEECH \cite{Zhang_WeNetSpeech_ICASSP2022}} & 
\multicolumn{1}{c|}{ASR} &  
\multicolumn{1}{c|}{[10kh labeled, 2.4kh weakly labeled, 10kh unlabeled]} &  
\multicolumn{1}{c|}{N/A} &  
\multicolumn{1}{c|}{16 kHz} &  
\multicolumn{1}{c}{Mandarin} \\
 
\midrule

\multicolumn{1}{l|}{DidiSpeech-1 \cite{Guo_DidiSpeech_ICASSP2021}} & 
\multicolumn{1}{c|}{VC, TTS, ASR} &  
\multicolumn{1}{c|}{572h} &  
\multicolumn{1}{c|}{4500 [2094, 2406]} &  
\multicolumn{1}{c|}{48 kHZ} &  
\multicolumn{1}{c}{Mandarin} \\
 
\midrule

\multicolumn{1}{l|}{DidiSpeech-2 \cite{Guo_DidiSpeech_ICASSP2021}} & 
\multicolumn{1}{c|}{VC, TTS, ASR} &  
\multicolumn{1}{c|}{227h} &  
\multicolumn{1}{c|}{1500 [797, 703]} &  
\multicolumn{1}{c|}{48 kHz} &  
\multicolumn{1}{c}{Mandarin} \\
 
\midrule

\multicolumn{1}{l|}{TITML-IDN \cite{Lestari_TITML_IDN_ISCPP2006}} &  
\multicolumn{1}{c|}{ASR} &  
\multicolumn{1}{c|}{14.5h} &  
\multicolumn{1}{c|}{20 [11, 9]} &  
\multicolumn{1}{c|}{16 kHz} &  
\multicolumn{1}{c}{Indonesian} \\

\midrule

\multicolumn{1}{l|}{INDspeech News LVCSR \cite{Sakti_LVCSR_TCAST2008}} &  
\multicolumn{1}{c|}{ASR} &  
\multicolumn{1}{c|}{43.35h} &  
\multicolumn{1}{c|}{400 [200, 200]} &
\multicolumn{1}{c|}{16 kHz} &  
\multicolumn{1}{c}{Indonesian} \\
 
\midrule

\multicolumn{1}{l|}{KhazakTTS 2\cite{Mussakhojayeva_KazakhTTS2_LRE2022}} & 
\multicolumn{1}{c|}{TTS} &  
\multicolumn{1}{c|}{271h} &  
\multicolumn{1}{c|}{5 [2, 3]} &  
\multicolumn{1}{c|}{22.05 kHz} &  
\multicolumn{1}{c}{Kazakh} \\

\midrule

\rowcolor{lightgray}
\multicolumn{6}{c}{Multilingual datasets} \\

\midrule

\multicolumn{1}{l|}{CommonVoice \cite{Ardila_CommonVoice_arxiv2019}} &  
\multicolumn{1}{c|}{ASR, LID} &  
\multicolumn{1}{c|}{2508h} &  
\multicolumn{1}{c|}{58250 [N/A, N/A]} &  
\multicolumn{1}{c|}{48 kHz} &  
\multicolumn{1}{c}{Multilingual} \\
 
\midrule

\multicolumn{1}{l|}{MLS \cite{Pratap_MLS_arxiv2020}} &  
\multicolumn{1}{c|}{ASR, TTS} &  
\multicolumn{1}{c|}{50.5kh} &  
\multicolumn{1}{c|}{6332 [3124, 3208]} &  
\multicolumn{1}{c|}{16 kHz} &  
\multicolumn{1}{c}{Multilingual} \\
 
\midrule

\multicolumn{1}{l|}{OpenSLR \cite{Sodimana_OpenSLR_SLTU2018}} &  
\multicolumn{1}{c|}{TTS} &  
\multicolumn{1}{c|}{25.4h} &  
\multicolumn{1}{c|}{139 [49, 90]} &  
\multicolumn{1}{c|}{22.05 kHz} &  
\multicolumn{1}{c}{Multilingual} \\
 
\midrule

\multicolumn{1}{l|}{Emilia \cite{He_Emilia_SLT2024}} &  
\multicolumn{1}{c|}{TTS} &  
\multicolumn{1}{c|}{101kh} &  
\multicolumn{1}{c|}{N/A} &  
\multicolumn{1}{c|}{24 kHz} &  
\multicolumn{1}{c}{Multilingual} \\
 
\midrule

\multicolumn{1}{l|}{VocCeleb1 \cite{Nagrani_VoxCeleb_arxiv2017}} & 
\multicolumn{1}{c|}{SI, SV} &  
\multicolumn{1}{c|}{352h} &  
\multicolumn{1}{c|}{1251 [690, 561]} &  
\multicolumn{1}{c|}{16 kHz} &  
\multicolumn{1}{c}{N/A} \\

\midrule

\multicolumn{1}{l|}{VocCeleb2 \cite{Chung_VoxCeleb2_arxiv2018}} & 
\multicolumn{1}{c|}{SI, SV} &  
\multicolumn{1}{c|}{2442h} &  
\multicolumn{1}{c|}{6112 [3761, 2351]} &  
\multicolumn{1}{c|}{N/A} &  
\multicolumn{1}{c}{N/A} \\

\midrule

\rowcolor{lightgray}
\multicolumn{6}{c}{Emotional datasets} \\

\midrule

\multicolumn{1}{l|}{IEMOCAP \cite{Busso_IEMOCAP_LRE2008}} &  
\multicolumn{1}{c|}{SER} &  
\multicolumn{1}{c|}{12h} &  
\multicolumn{1}{c|}{10 [5, 5]} &  
\multicolumn{1}{c|}{48kHz} &  
\multicolumn{1}{c}{English} \\

\midrule

\multicolumn{1}{l|}{ESD \cite{Zhou_ESD_SpeechCom2022}} &  
\multicolumn{1}{c|}{EVC} &  
\multicolumn{1}{c|}{29h} &  
\multicolumn{1}{c|}{20 [N/A, N/A]} &  
\multicolumn{1}{c|}{N/A} &  
\multicolumn{1}{c}{English, Mandarin} \\

\bottomrule

\end{tabular}}
\label{SpeechRelatedDatasets}
\end{table*}

\begin{table*}[t]
\centering
\caption{Evaluation Metrics, where each variable will correspond to either the reference X or the prediction $X^{'}$.}
\renewcommand{\arraystretch}{1.7}

\resizebox{18cm}{!}{
\begin{tabular}{cccc}
\toprule
\textbf{Metric} & \textbf{Formula/Scale} & \textbf{Focus} & \textbf{Task} \\ \midrule

\rowcolor{lightgray}
\multicolumn{4}{c}{Objective} \\
\midrule

\multicolumn{1}{l|}{MCD \cite{Kubicheck_MCD_PRCCCSP1993} $\downarrow$} &  
\multicolumn{1}{c|}{\(\frac{1}{T} \sum_{t=0}^{T-1} \sqrt{\sum_{k=1}^{K}(c_{t,k} - c^{'}_{t,k})^{2}}\)} &  
\multicolumn{1}{c|}{Speech Quality} &
\multicolumn{1}{l}{Measures spectrum similarity, where $c_{t,k}$ is the k-th MFCC of t-th time frame from audio waveform.} \\

\midrule

\multicolumn{1}{l|}{CER/WER $\downarrow$} &  
\multicolumn{1}{c|}{\(\frac{\text{Insertions} + \text{Deletions} + \text{Substitutions}}{\text{All characters/Words}}\)} &  
\multicolumn{1}{c|}{Speech Quality} &
\multicolumn{1}{l}{Measures CER/WER of transcribed speech using a pre-trained ASR and the reference text.} \\
 
\midrule

\multicolumn{1}{l|}{SECS $\uparrow$} &  
\multicolumn{1}{c|}{\(\frac{Embeddings \cdot Embeddings^{'}}{\|Embeddings\| \times \|Embeddings^{'}\|}\)} &  
\multicolumn{1}{c|}{Speaker Similarity} &
\multicolumn{1}{l}{Measures the cosine similarity between the speaker's generated embedding and the ground truth.} \\
 
\midrule

\multicolumn{1}{l|}{SV-EER \cite{Hansen_SV_EER_SPM2015} $\downarrow$} &  
\multicolumn{1}{c|}{\(\frac{\text{False Accept (FA) Errors}}{\text{Imposter Attempts}}\) = \(\frac{\text{False Reject (FR) Errors}}{\text{Legitimate Attempts}}\)} &  
\multicolumn{1}{c|}{Speaker Similarity} &
\multicolumn{1}{l}{Measures the performance of speaker verification systems, where False-Acceptance Rate (FAR) = False-Rejection Rate (FRR).} \\
 
\midrule

\multicolumn{1}{l|}{GPE \cite{Nakatani_GPE_SpeechCom2008} $\downarrow$} &  
\multicolumn{1}{c|}{\(\frac{\sum_{t} [|p_{t} - p^{'}_{t}| > 0.2p_{t}]1[v_{t}]1[v^{'}_{t}]}{\sum_{t}1[v_{t}]1[v^{'}_{t}]}\)} &  
\multicolumn{1}{c|}{Speech Expression} &
\multicolumn{1}{l}{Voice Frame percentage with at least 20\% from reference. $p_{t}$ is the pitch signal from audio, and $v_{t}$ is voicing decision.} \\
 
\midrule

\multicolumn{1}{l|}{VDE \cite{Nakatani_GPE_SpeechCom2008} $\downarrow$} &  
\multicolumn{1}{c|}{\(\frac{\sum_{t=0}^{T-1} 1 [v_{t} \neq v^{'}_{t}]}{T}\)} &  
\multicolumn{1}{c|}{Speech Expression} &
\multicolumn{1}{l}{Frame percentage with incorrect voicing decisions. $v_{t}$ is voicing decisions audio, where T is the total frames.} \\
 
\midrule

\multicolumn{1}{l|}{FFE \cite{Chu_FFE_ICASSP2009} $\downarrow$} &  
\multicolumn{1}{c|}{\(\frac{\sum_{t}^{T-1} [|p_{t} - p^{'}_{t}| > 0.2p_{t}]1[v_{t}]1[v^{'}_{t}] + 1[v_{t} \neq v_{t}^{'}]}{T}\)} &  
\multicolumn{1}{c|}{Speech Expression} &
\multicolumn{1}{l}{Percentage of frames where an error either by VDE or GPE occurs.} \\
 
\midrule

\multicolumn{1}{l|}{RTF $\downarrow$} &  
\multicolumn{1}{c|}{NA} &  
\multicolumn{1}{c|}{Inference Speed} &
\multicolumn{1}{l}{The time in seconds that the device requires to synthesize one second of the waveform.} \\

\midrule

\rowcolor{lightgray}
\multicolumn{4}{c}{Subjective} \\
\midrule

\multicolumn{1}{l|}{MOS $\uparrow$} &  
\multicolumn{1}{c|}{5-point scale} & 
\multicolumn{1}{c|}{NA} &
\multicolumn{1}{l}{Average rating across evaluators following a point-scale. This is used to evaluate naturalness, similarity, quality, and intelligibility.} \\
 
\midrule

\multicolumn{1}{l|}{CMOS $\uparrow$} &  
\multicolumn{1}{c|}{7-point scale} &  
\multicolumn{1}{c|}{NA} &
\multicolumn{1}{l}{The evaluator rates preference on a 7-point scale (-3, 3), where positive values indicate improvement while negative indicates a decline.} \\
 
\midrule

\multicolumn{1}{l|}{MUSHRA \cite{Series_MUSHRA_ITU2014} $\uparrow$} &  
\multicolumn{1}{c|}{100-point scale} &  
\multicolumn{1}{c|}{NA} &
\multicolumn{1}{l}{Evaluator rates speech, including reference, generated, hidden, and anchor signals with access to all signals for consistent comparison.} \\
 
\midrule

\multicolumn{1}{l|}{AB preference test $\uparrow$} &  
\multicolumn{1}{c|}{Model A \%: Model B \%: No Preference \%} &  
\multicolumn{1}{c|}{NA} &
\multicolumn{1}{l}{Evaluator compares synthesized speech of two methods. If more than two methods, this method can be called ABX preference test.} \\

\bottomrule

\end{tabular}}
\label{Evaluation_Metrics}
\end{table*}

\section{Datasets \& Evaluation Metrics}
\label{Section5}

Open-source speech datasets, summarized in Table \ref{SpeechRelatedDatasets}, are vital for advancing monolingual and multilingual voice cloning algorithms. For example, speaker verification can improve speaker similarity, while ASR pre-trained models are used to test the intelligibility of the synthesized speech.

Evaluating voice cloning algorithms is crucial for an in-depth comparison focusing on various TTS aspects and the evaluation metrics are summarized in Table \ref{Evaluation_Metrics}. Speech naturalness measures how natural the voice sounds, measured subjectively using the Mean Opinion Score (MOS). Speech quality focuses on the clarity and intelligibility of the speech, evaluated using Mel-Cepstral Distortion (MCD), Character Error Rate (CER), and Word Error Rate (WER), in which the latter two depend on the pre-trained ASR system used for evaluation. Speech similarity focuses on the resemblance between the generated voice and the target speaker, evaluated using Speaker Embedding Cosine Similarity (SECS), which calculates the cosine similarity between the embeddings of the generated speech and the reference audio. Although this metric is objective, there is still some subjectivity due to the speaker encoder choice. Speech expression evaluates the style and emotions of the generated speech using prosody-related metrics such as Gross Pitch Error (GPE), Voicing Decision Error (VDE), and F0 Frame Error (FFE). Finally, Generation speed can be quantified using the real-time factor (RTF), indicating the time required to generate one second of audio.

\section{Applications \& Misuse}
\label{Section6}

\textbf{Entertainment \& Content Creation:}
Voice cloning is highly valuable in the entertainment and content creation industry. For instance, automating voice-over for animated movies saves time and allows more focus on film design \cite{Neekhara_ACML2021}. Moreover, it can speed up the movie dubbing process, expanding its reach to international audiences while preserving voice characteristics and emotions \cite{Li_ZSE_VITS_Electronics2023}. Content creators can benefit by having the option to edit audio narration to insert words in case of misspellings during the initial video \cite{Tang_InterSpeech2021}.

\textbf{Personal Assistants:}
Personalized assistants can benefit from expressive and realistic sounds as they improve the user experience. For example, the ability to generate speech with an appropriate emotional tone can ensure the user is engaged and comforted with the expressiveness of the AI-generated speech. Cerence introduced "My Car, My Voice" to let users create custom voices for their in-car assistants \cite{Cerence_mycarmyvoice_2024}.

\textbf{Advertisement:}
Voice cloning offers various possibilities to transform the advertisement industry \cite{Kadam_Survey_EAI2021}. It allows the creation of a brand voice tailored to different target audiences. This also includes multilingual voice cloning, where you can maintain the speaker's characteristics to reach a higher audience.

\textbf{Accessibility:}
Few-shot and zero-shot voice cloning can provide value toward restoring the communication ability of people who lost their voice \cite{Jia_NIPS2018}. Dysphonia is a speech disorder leading to unclear and difficult-to-understand speech, affecting their social life and mental well-being. Voice cloning can assist in such cases as it can improve their speech intelligibility while maintaining voice characteristics \cite{Wadoux_TSD2023}.

\textbf{Misuse:}
Voice cloning algorithm misuse has been occurring lately at an alarming rate as less data is required to replicate someone's voice. It is essential to understand that anyone can be a victim of an impersonation or financial scam. Hutiri et al. \cite{Hutiri_ACM2024} presented a speech generation harm taxonomy, discussing various aspects of the non-ethical usage of voice cloning algorithms. For example, using people's voices by training and releasing their AI models can lead to identity theft, data laundering, impersonation, and coercion. Furthermore, voice cloning can lead to consumer fraud and financial losses. In August of 2019, a company's CEO was tricked by an audio deepfake to perform a fraudulent transfer of \$243,000 \cite{Damiani_Forbes2019}, proving the importance of understanding its potential misuse and developing detection algorithms. 

\section{Future Research Directions}
\label{Section7}

\textbf{Lack of TTS Benchmark:}
Researchers are focused on enhancing monolingual and multilingual voice cloning. However, objective comparisons are tricky due to testing set variation across papers, making it difficult to grasp the status of voice cloning. Establishing standardized methods for evaluation is crucial for further voice cloning development.

\textbf{Conversation-Style Speech Synthesis:}
TTS research focuses mainly on synthesizing reading-style speech, disregarding casual tones, which can be helpful in conversational style generation. There are very few datasets aimed at addressing this type of task, so exploring conversational TTS could be an exciting area of research. AdaSpeech 3 \cite{AdaSpeech3_InterSpeech2021}  focuses on spontaneous speech synthesis, mimicking how people interact in conversations or a podcast setting. 

\textbf{Fine-grained Disentanglement:}
Representation disentanglement is a crucial technique for achieving fine-grained control over TTS systems. Even though most work concentrates on content-timbre disentanglement, there is a critical need for enhanced control over prosodic features for manipulating voice expression effectively. The ability to replicate emotions and expressions can be a valuable asset for many HCI applications.

\section{Conclusion}
Voice cloning research has increased rapidly, driven by the industrial interest in digitization and digital twins. This survey systematically examines the advancements made in synthetic voice generation, specifically voice cloning. We categorize the methods into four main areas: speaker adaptation, few-shot voice cloning, zero-shot voice cloning, and multilingual voice cloning. We then classify those works based on their contributions and the challenges they tackle to enhance voice cloning performance. We also explore speech-related datasets and the evaluation metrics used to compare voice cloning algorithms. Understanding these methods is essential for enhancing their performance and developing algorithms for their detection, thereby mitigating the risks of audio deepfake generation.

\section{Biography Section}

\begin{IEEEbiographynophoto}{Hussam Azzuni} received an MSc degree in Computer Vision from MBZUAI, UAE. He is passionate about developing Human-AI interaction algorithms and 3D reconstruction methodologies for AR/VR immersive experiences. 
\end{IEEEbiographynophoto}

\begin{IEEEbiographynophoto} {Abdulmotaleb El Saddik} (Fellow, IEEE) is a Distinguished University Professor specializing in intelligent multimedia computing and digital twin technologies. His research integrates various technologies, including AI, IoT, AR/VR, haptics, and 5G, to enhance real-time interactions in the metaverse. He served as Editor-in-Chief, authored numerous publications and books, chaired conferences/workshops, and supervised over 150 researchers.
\end{IEEEbiographynophoto}

\vfill

\end{document}